\documentclass[journal]{IEEEtran}

\usepackage{fancyhdr}
\usepackage{url}
\usepackage[dvipsnames]{xcolor}
\usepackage{graphicx,subfigure}
\usepackage{epsfig,cite}
\usepackage{color}
\usepackage{amssymb}
\usepackage{algorithm} 
\usepackage{algorithmic}
\usepackage{upgreek}
\usepackage{listings}
\usepackage{eqnarray}
\usepackage{tikz}
\usetikzlibrary{automata,arrows,positioning,calc}
\usepackage{eqnarray,amsmath}
\usepackage{float}
\usepackage{caption}
\usepackage{multirow}
\usepackage{rotating}
\usepackage{makecell}
\usepackage{amsmath}
\usepackage[pagewise,left]{lineno}

\newcommand{\quotes}[1]{``#1''}

\newcount\cyclecount \cyclecount=-1
\newcount\ind \ind=-1

\renewcommand\headheight{24pt}
\definecolor{dkgreen}{rgb}{0,0.6,0}
\definecolor{gray}{rgb}{0.5,0.5,0.5}
\definecolor{mauve}{rgb}{0.58,0,0.82}

\ifCLASSINFOpdf
\else
\fi
\begin{document}
%
\title{Modeling the Propagation of Trojan Malware \\ in Online Social Networks}
%
%
%

\author{Mohamamd~Reza~Faghani,~\IEEEmembership{Member,~IEEE,}
        and~Uyen~Trang~Nguyen,~\IEEEmembership{Member,~IEEE,}
\thanks{The authors are with the Department of Electrical Engineering and Computer Science, York University, Toronto, Ontario, M3J 1P3, Canada, Email: \{faghani,utn\}@cse.yorku.ca.}}

%
%

\markboth{IEEE Transactions on Dependable and Secure Computing ,~Vol.~15, June~2017}%
{Faghani \MakeLowercase{\textit{et al.}}: Modeling the Propagation of Trojan Malware in Online Social Networks}
%



\maketitle

\begin{abstract}
The popularity and widespread usage of online social networks (OSN) have attracted cyber criminals who have used OSNs as a platform to spread malware.  Among different types of malware in OSNs, Trojan is the most popular type with hundreds of attacks on OSN users in the past few years.  Trojans infecting a user's computer have the ability to steal confidential information, install ransomware and infect other computers in the network.  Therefore, it is important to understand propagation dynamics of Trojans in OSNs in order to detect, contain and remove them as early as possible.  In this article, we present an analytical model to study propagation characteristics of Trojans and factors that impact their propagation in an online social network.  The proposed model assumes all the topological characteristics of real online social networks. Moreover, the model takes into account attacking trends of modern Trojans, the role of anti-virus (AV) products, and security practices of OSN users and AV software providers. By taking into account these factors, the proposed model can accurately and realistically estimate the infection rate caused by a Trojan malware in an OSN as well as the recovery rate of the user population.	

\end{abstract}


%
\IEEEpeerreviewmaketitle

%
%
%
%

\section{Introduction}
\label{intro}
\IEEEPARstart{O}{line} social networking are amongst the most popular services offered through the World Wide Web. Online social networks (OSNs) such as Facebook, Twitter and MySpace have provided hundreds of millions of people with a means to connect and communicate with their friends, families and colleagues around the world. For instance, Facebook is the second most visited website in the world according to a recent ranking by Alexa \cite{alexa},  only after Google. 

The popularity and wide spread usage of OSNs have attracted hackers and cyber criminals to use OSNs as an attack platform to spread malware \cite{grossman,facebook-virus}.   A successful attack using malware in an OSN can lead to tens of millions of OSN accounts being compromised and users' computers being infected.   Cyber criminals can mount massive denial of service attacks against Internet infrastructures or systems using compromised accounts and computers.  They can steal users' sensitive information for fraudulent activities.  Compromised OSN accounts can also be used to spread misinformation to bias public opinions \cite{opin}, or even to influence automatic trading algorithms that rely on public opinions \cite{designsocialbot}.  (Automatic trading algorithms place buy/sell stock orders on behalf of human investors.)


\subsection{Overview of OSN Malware}
There are two major types of malware that target online social network users: : cross-site scripting worm and Trojan:

\begin{itemize}
	\item Cross-site scripting (XSS) worms: These are passive malware that exploits vulnerabilities in web applications to propagate themselves without any user intervention. 
	\item Trojans: A Trojan is a type of malware that is often disguised as legitimate software. Users are typically tricked by some form of social engineering into opening them (and thus loading and executing Trojans on their systems).  Trojans are the most common method used to launch attacks against OSNs users, who are tricked into visiting malicious websites and subsequently downloading malware disguised as legitimate software (e.g., Adobe Flash player). There are many variants of Trojans operating in OSNs, including clickjacking worms \cite{cj} and extension-based malware\cite{extmalware}. 
\end{itemize}

Compared with XSS worm, Trojan is the more popular type of malware targeting OSN users. Over the past few years, Facebook users have experienced hundreds of separate Trojan malware attacks \cite{kilim1,kasper,koobface}. For instance, the first variant of an OSN Trojan browser extension called Kilim appeared in November 2014 \cite{kilim1}.  From November 2014 to November 2016, almost 600 variants of Kilim  were discovered \cite{kilim2}. In most cases, a Trojan disguises itself as a legitimate software. For instance in two major Trojan attacks on Facebook, the Trojan posed itself as an Adobe Flash player update \cite{koobface, kasper}. In a more recent attack discovered in 2015 \cite{kasper}, a message enticed the victims to click on a link that redirected them to a third-party website unaffiliated with Facebook where they were prompted to download what was claimed to be an update of the Adobe Flash player. If they downloaded and executed the file, they would infect their computers with a Trojan malware.

Trojans installed on a user's computer have the ability to access contents on the compromised system, including social network contents, credit card information, and login credentials. It can even spread itself further by infecting other systems on the same network. Such Trojans have the ability to form a \textit{botnet} to open up channels for attackers to send further payloads such as ransomware. Such a Trojan is a variant of Locky ransomware discovered in November 2016 \cite{checkpoint}, which  was delivered via JPEG and SVG files via Facebook Messenger.

\subsection{Motivations}

Given the popularity of and potential damages inflicted by Trojans, it is important to understand their propagation dynamics in OSNs in order to detect, contain and remove them as early as possible.   Therefore, our objective is to model and study the propagation of Trojans in social networks such as Facebook, LinkedIn and Orkut. (These networks can be represented by undirected graphs, in which each vertex represents a user, and each edge represents the mutual relationship between the two users denoted by the two end vertices.)

The topic of Trojan propagation in OSNs has only been studied recently.   Most of these studies are based on simulations \cite{facebook-virus, dyn, ftrojan, fworm}.  Thomas et al. \cite{koobface} traced the activities of Koobface, a Trojan that targeted OSN users, for one month to study its propagation characteristics.

There exist few works on the topic of modeling propagations of malware in OSNs.  Faghani and his collaborators  \cite{fxssj, fxss} modeled the propagation of XSS worms in OSNs.   Sanzgiri et al. \cite{twitjack} modeled the propagation of Trojans in the social network Twitter where most relationships are one-directional (follower-followee), unlike mutual relationships in Facebook or LinkedIn networks.

There exist works that model propagations of worms and malware (not necessarily Trojans) in other types of networks such as people, email and cellular phones.  Many of these models \cite{boguna,moreno,moreno2,pastor} assumed that each user is directly connected to every other user in the same network (also known as ``homogeneous mixing'').  This assumption does not hold true for a real-world OSN such as Facebook where each user is directly connected to only his/her friends.  As a result, the ``homogeneous mixing'' assumption  may lead to an over-estimation of the infection rate in a real OSN \cite{wen,email}.  Cheng et al. \cite{smartchen} proposed a propagation model for malware that targets multimedia messaging service (MMS) and bluetooth devices.  Chen and Ji \cite{chen} and Chen et al. \cite{sii1} modeled the spreading of scanning worms\footnote{Scanning worms scan targets, such as computers, routers, etc. for exploitable vulnerabilities in order to deliver malicious payloads via these exploits.} in computer networks.  Zou et al. \cite{email}  and Komnios et al.  \cite{komin} studied the propagation of email worms.    Wen et al. \cite{wen} also modeled the propagation of malware in email networks and in semi-directed networks represented by mixed graphs (i.e., a subset of edges are directed while the others are undirected).

Besides the network topology, there are several other factors that affect the propagation of Trojans in social networks.  For example, anti-virus (AV) products play an important role in protecting users against malware and thus slowing down their  propagation.  On the other hand, malware programs have recently become more and more sophisticated and able to evade detection by AV products.  A recent test of AV products on the market performed by Virus Bulletin in 2016 \cite{avnews} shows that the rate of detection of unknown malware samples is around 67\% to 70\%.  (The same article suggests that this number will get smaller over time as malware  becomes more sophisticated.)   In response to novel malware, AV vendors need to update their products and provide patches against them.  It takes time to understand the working mechanism of a new malware and come up with solutions against it.   The faster AV vendors release updates/patches, the higher the chance of stopping the malware.  Another factor that impacts the propagation of recent Trojans is their ability to prevent users from accessing websites of AV vendors so that the users cannot download updates/patches.  There have been several instances of such malware \cite{kasper,koobface, extmalware, conficker, sophoslab, ghost} including those targeting Facebook users such as Magnet \cite{kasper}, Koobface \cite{koobface}, and extension-based malware \cite{extmalware}.  In response to this new ability of Trojans, infected users have reached out to their OSN friends, asking for clean-up solutions to remove the malware from their systems, as in the case of a large-scale attack on Facebook caused by a malware named Magnet \cite{kasper}.  All the above factors have an impact on propagation dynamics of Trojans in OSNs; however, none of previous works on modeling worms/malware in OSNs considered these factors.

\subsection{Contributions}

Having identified gaps in existing research, we propose an analytical model that

\begin{itemize}
	
	\item considers characteristics of modern Trojans (e.g, malware blocking users' access to AV provider websites), security practices (e.g., users installing AV products on their computers, AV manufacturers gradually releasing updates/patches against a newly propagating malware), and user behaviour (e.g., seeking assistance from OSN friends to clean up infected computers).  None of previous works on modeling worms/malware in OSNs considered the above factors.
	
	\item assumes the topological characteristics of real-world social networks, namely,  low average shortest distance, power-law distribution of node degrees and high clustering coefficient \cite{dekker, holme, mislov}. In this article, we consider OSNs that are represented by undirected graphs such as Facebook, Linked and Orkut. To the best of our knowledge, our work is the first that models Trojan propagation in such networks.  (In the future we will extend the model to OSNs represented by directed graphs such as Twitter.) 
	
	\item is validated using a real-world social network graph, a Facebook sub-graph constructed by McAuley and Leskovec \cite{snap} that possess all the characteristics of online social networks as mentioned above. In all experiments we studied,  numerical results obtained from the model closely match simulation results, demonstrating the accuracy of the model.
	
	\item has low computational complexity while being accurate and taking into account a wide range of influencing factors discussed above. In particular, the computational complexity is $O(E)$, where $E$ is the number of edges in the network graph.
	
\end{itemize}

In the future, Trojan malware may become even more sophisticated with more attacking vectors, which we may not be able to anticipate presently.  By taking into account attacking trends of  modern Trojans and security practices, our proposed model can be extended for future modeling of OSN Trojans that have more sophisticated attacking mechanisms. 

The remainder of this article is organized as follows. In Section \ref{characs}, we describe the topological characteristics of OSNs. In Section \ref{experiment}, we discuss the propagation mechanism of Trojan malware in OSNs. In Sections \ref{modeldes} and \ref{model_detail}, we describe the proposed analytical model. We validate the model in Section \ref{modelperf}. We discuss related work in Section \ref{relworks} and finally we conclude the article in Section \ref{concl}.

\section{Characteristics of Online Social Networks}
\label{characs}
An OSN can be represented by an equivalent graph in which each vertex (or node) represents a person, and a link between two vertices indicates a relationship (friendship) between the two respective persons.   
There exist many OSNs in which the relationship (friendship) between two users is mutual (e.g., Facebook, LinkedIn, Orkut), and thus, they can be represented by undirected graphs.

Online social networks have distinct characteristics that distinguish them from other types of networks such as computer networks. Following are the characteristics of OSNs \cite{dekker, holme, mislov}:

\begin{enumerate}
	\item 
	An OSN typically has a low average shortest path distance, approximately equal to  $\log(s)/\log(d)$, where $s$ is the number of vertices (people), and $d$ is the average vertex degree of the equivalent graph \cite{holme}.
	
	\item 
	Online social networks typically show a high clustering property, or high local transitivity.  That is, if person $A$ knows $B$ and $C$, then $B$ and $C$ are likely to know each other. Thus $A$, $B$ and $C$ form a friendship triangle. Let $k$ denote the degree of a vertex $v$. Then the number of all possible triangles originated from vertex $v$ is $k(k-1)/2$.  Let $f$ denote the number of friendship triangles of a vertex $v$ in a social network graph. Then the clustering coefficient $C(v)$ of vertex $v$ is defined as $C(v) = 2f/(k(k-1))$. The clustering coefficient of a graph is the average of the clustering coefficients of all of its vertices. In a real OSN, the average clustering coefficient is about 0.1 to 0.7 \cite{holme, dekker}.
	
	\item 
	Node degrees of a social network graph tend to be power-law distributed.  The node degree of a power-law topology is a right-skewed distribution with a power-law Complementary Cumulative Density Function (CCDF) of $F(k)\propto k^{-\alpha}$, which is linear on a logarithmic scale. The power law distribution states that the probability for a node $v$ to have a degree $k$ is $P(k)\propto k^{-\alpha}$, where $\alpha$ is the power-law exponent \cite{nman}.
\end{enumerate}

To validate our proposed model and run simulations, we used a real-world graph that possesses all the characteristics of a social network.  The graph was derived from a Facebook data set provided by McAuley and Leskovec \cite{snap} as part of the Stanford Network Analysis Project (SNAP).  This data set was built in 2012 using a Facebook application that conducted a survey to collect users' information such as their friend lists and  shared information on their profiles \cite{snap}. The resulting graph has all the OSN characteristics discussed above. Table \ref{OSN_graphs} summarizes the characteristics of this social network graph.

\begin{table}[t]
	\caption{Parameters of the Facebook sub-graph}
	\centering
	\label{OSN_graphs}
	\begin{tabular}{llllll}
		\hline\noalign{\smallskip}
		Parameter       & OSN   \\ 
		\noalign{\smallskip}\hline\noalign{\smallskip}
		Number of vertices (people)	& 4,039 	\\ 
		Number of edges 		& 88,234 	\\ 
		Average clustering coefficient	& 0.60 	\\ 
		Average shortest path length	&  3.7  	\\ 
		Network diameter		& 8	 	\\
		Maximum node degree		& 45 \\  
		Average node degree $d$		& 43.69  \\ 
		$\log(N)/\log(d)$		& 1.6		\\ 
		\noalign{\smallskip}\hline
	\end{tabular}
\end{table}

\section{Propagation Mechanism of Trojan Malware in Online Social Networks}
\label{experiment}
In this section, we describe the process used by one or more real-world Trojan malware (e.g., Koobface \cite{koobface}, Magnet \cite{kasper} and generic extension based malware \cite{extmalware} such as Kilim \cite{kilim1, kilim2} to attack OSN users and propagate in a network. Such a process consists of three stages:

\begin{enumerate}
	\item In the first stage, the malware developer creates one or more fake profiles and infiltrates them into the social network. The purpose of these fake profiles is to make friends with as many real OSN users as possible.  Infiltration has been shown to be an effective technique for disseminating malicious content in  OSNs such as  Facebook \cite{designsocialbot}. 
	
	\item In the second stage, the malware developer uses social engineering techniques to create eye-catching web links that trick users into clicking on them. The web links, which are posted on the fake users' walls, lead unsuspecting users to a web page that contains malicious content.   A user simply needs to visit or ``drive by'' that web page, and the malicious code can be downloaded in the background and executed on the user's computer without his/her knowledge.  This type of attack is called \textit{drive-by download} and is caused by vulnerabilities in browsers, apps or operating systems that are out of date and have security flaws~\cite{dis}.  
	
	When security flaws are absent, malware creators resort to social engineering techniques to get assistance from users to activate the malicious code. For instance,  after a user lands on the malicious web page, he/she is asked to click on a button to download a software  (e.g., an updated version of the Adobe Flash player) or to play a video.  If the user clicks on the button, he/she is actually downloading and executing a malware.  
	
	In either case, drive-by or user-assisted download, the user's computer is infected.  The computer  can then be controlled by the attacker(s) to perform other malicious activities such as stealing confidential information stored on the computer,  attacking other computers on the same network, or mounting denial of service attacks against vulnerable websites.
	
	\item In the third stage, after a user $u$ is infected, the malware also posts the eye-catching web link(s) on the user's wall  (i.e., via newsfeed), to ``recruit'' his/her friends.  If a friend of $u$ clicks on the link(s) and, as a result, \textit{unknowingly} executes the malware, the friend's computer and profile will become infected as described in stage 2 and the propagation cycle continues with his/her own friends. 
\end{enumerate}

\begin{figure*}[t]
	\centering
	\includegraphics[width=5in]{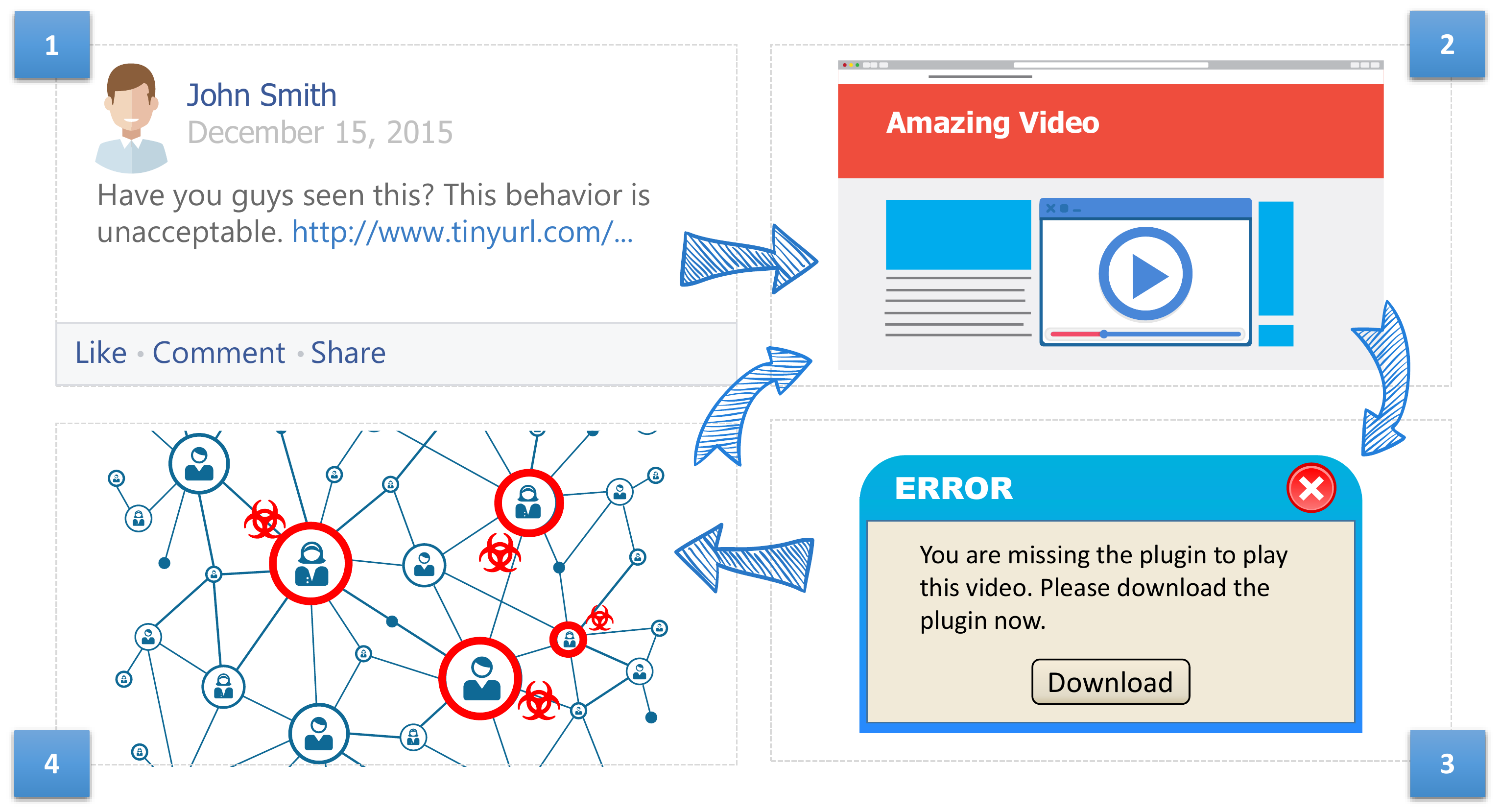}
	\caption{Example of Trojan malware propagation in online social networks}
	\label{cycle}
\end{figure*}

The above process is illustrated by the diagram shown in Fig. \ref{cycle}.  In this example, the malware posts a malicious link on an infected user's wall, enticing the infected user's friends to watch a video (step 1). Friends of the infected user who follow the malicious link will land on a web page that hosts an object that looks similar to a video player (step 2). After clicking the \quotes{play} button, the friend receives a notification that he/she requires a software (or a \quotes{plugin}) in order to watch the video (step 3). Users who download and execute the \quotes{plugin} are actually and unknowingly executing the malware and become infected (step 4). The malware will then post the malicious link on the walls of the newly infected users to lure their friends to the web page that hosts the fake video player, and the propagation cycle continues with more new victims. Note that the wall concept may not exist in some social networks.  In this case, we can assume, without loss of generality, that the malware sends the malicious link directly to an infected user's friends via the OSN private messaging system.

\section{Overview of the Proposed Malware Propagation Model}
\label{modeldes}

In this section, we present an overview of the proposed model and underlying assumptions.

\subsection{Modeling Approach}
In the literature, epidemic models can be classified into two main categories: SIS (susceptible-infected-susceptible) and SIR (susceptible-infected-recovered).  SIS models~\cite{pastor,p2p,sis1,sis2} assume that an infected user, after being disinfected, will become susceptible again and thus may be re-infected by the same disease (malware).  SIR models~\cite{moreno,moreno2,boguna,chen,wen}, on the other hand, assume that an infected user, after being disinfected (having recovered or becoming immune), will not be re-infected again by the same disease (malware).

The SIS approach is not suitable for modeling Trojans in online social networks, because it does not support the immune state.  In practice, users who have anti-virus software installed may be immune against a particular malware.

Our proposed model is based on the SIR approach.  Previous SIR models suffer from one or more of the following drawbacks:
\begin{itemize}
	\item Many models \cite{pastor, moreno, moreno2, boguna} assume homogeneous mixing as mentioned earlier, and thus overestimate the infection rate in a real OSN.
	\item Some models \cite{chen, smartchen} assume that users check newly arriving messages at every time unit, and all users have the same message checking time (usually one time interval).  These models do not consider the temporal dynamics of user activities.
	\item Some models incur high computational complexity, such as $O(E^2)$ \cite{wen}, where $E$ is the number of edges (friendships) of the social network graph.  
\end{itemize}

Our proposed model is a spatial-temporal SIR model that takes into account the  topological characteristics of real-world social networks and temporal dynamics of user activities.  Furthermore, it considers characteristics of modern Trojans (e.g, malware blocking users' access to AV provider websites), security practices (e.g., users installing AV products on their computers, AV manufacturers gradually releasing updates/patches against a newly propagating malware), and user behaviour (e.g., seeking assistance from OSN friends to clean up infected computers).  None of previous works on modeling worms/malware in OSNs considered the above factors.  The proposed model has low computational complexity,  in the order of $O(E)$.

\subsection{User States}
We assume that all the users are vulnerable to the new malware \textit{M} when it first appears in the social network, i.e., at $t = 0$. That is, all users are in the  \textit{susceptible} state at time $t = 0$. As  time passes, susceptible users may stay susceptible, or transition to the \textit{immune} state thanks to a defensive mechanism such as an antivirus program against malware  $M$, or become \textit{infected} by the malware after clicking on the malicious link.  \textit{Infected} users can \textit{recover} by finding clean-up solutions to remove the malware from their systems, or may stay infected. If users have recovered from an infection or become immune, they are  no longer vulnerable to malware \textit{M} and thus will no longer be infected by it.  

Therefore,  at a specific point in time, each node (user) in the network can be in one of the following four different states with respect to a particular malware \textit{M}: \textit{susceptible}, \textit{infected}, \textit{recovered} and \textit{immune}.

\begin{itemize}
	\item A \textit{susceptible} node is a node that is vulnerable to malware infection but otherwise \quotes{healthy}. 
	\item An \textit{infected} node is a node that became infected and may potentially infect other nodes. 
	\item A \textit{recovered} node is a node that was infected but the user was able to find solutions to remove the malware from his computer and profile; the node is thus no longer infected or susceptible to the malware \textit{M}.
	\item An \textit{immune} node is a node that is unable to become infected thanks to a defensive mechanism existing on the system (e.g., having an antivirus program able to detect and block the malware \textit{M}). A node may also be immune to the malware \textit{M} because the user's operating system is not targeted by the malware.  For example, the malware exploits a vulnerability in and  attacks only Windows systems, but not Linux or Mac systems.
\end{itemize}

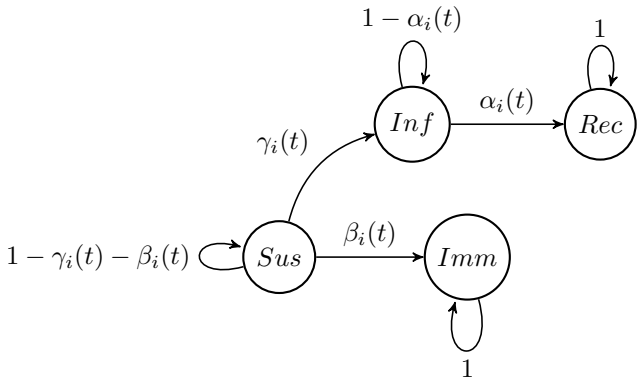
\begin{figure}[h]
	\centering
	\begin{tikzpicture}[->, >=stealth', auto, semithick, node distance=2.5cm]
	\tikzstyle{every state}=[fill=white,draw=black,thick,text=black,scale=1]
	\node[state]    (0)                     {$Sus$};
	\node[state]    (1)[above right of=0]   {$Inf$};
	\node[state]    (2)[right of=1]   {$Rec$};
	
	\node[state]    (3)[right of=0]   {$Imm$};
	\path
	(0) edge[loop left]     node{$1-\gamma_i(t)-\beta_i(t)$}         (0)
	edge[bend left]     node{$\gamma_i(t)$}     (1)
	edge[left, above]      node{$\beta_i(t)$}      (3)
	(1) edge [left,above]                node{$\alpha_i(t)$}           (2)
	edge[loop above]     node{$1-\alpha_i(t)$}         (1)
	(2) edge[loop above]    node{$1$}     (2)
	(3) edge[loop below]    node{$1$}     (3)
	;
	\end{tikzpicture}
	\captionof{figure}{State transition diagram of node $i$}
	\label{pm}
\end{figure}

At time $t = 0$, all users are in the \textit{susceptible} state.  At each time unit $t > 0$, a user $i$ may move from one state to another, as shown in  Figure \ref{pm}, which depicts the state transition diagram of a user.  The definitions of the transitional probabilities $\gamma_i(t)$, $\beta_i(t)$ and $\alpha_i(t)$ are provided in Table~\ref{sum}; their computations are discussed in Sections \ref{stoi}, \ref{sus2immune} and ~\ref{transinfimm}, respectively.

To model the user states, we define random variable $X_i(t)$ to denote the state of node $i$ at each time unit $t$, as follows: 

\begin{equation}
	\label{eq0}
	X_i(t) = \begin{cases}
		Sus, & \text{if node $i$ is susceptible at time $t$;}\\
		Inf, &  \text{if node $i$ is infected at time $t$;} \\
		Rec, &  \text{if node $i$ is recovered at time $t$;} \\
		Imm &  \text{if node $i$ is immune at time $t$} \\
	\end{cases}
\end{equation}

\subsection{Temporal Dynamics}

We consider the temporal dynamics of user activities by defining $\tau_i$ as the message checking time period of user $i$. That is, user $i$ visits the social network and check new posts or messages every $\tau_i$ time units.   This definition of user message checking time is common in previous works~\cite{wen, email,chen}.
According to this definition, at each time unit $t$, all the users whose $\tau_i$ values satisfying ($t\ mod \ \tau_i = 0$) visit the OSN and check their messages.   A user $i$'s message checking time is defined formally using a discrete random variable $P(visit_i(t)=true)$ as follows:
	
	\begin{equation}
		\label{timeq}
		P(visit_i(t)=true) = \begin{cases}
			1, & \text{if $t\ mod \ \tau_i =0$;}\\
			0 &  \text{otherwise} \\
		\end{cases}
	\end{equation}

For instance, if $\tau_i = 10$, user $i$ would visit the OSN and check messages at time $t = 10, 20, 30, \dots$.  Another user $j$ may vist the OSN more often than user $i$ and has $\tau_j = 4$.  User $j$ would visit the OSN and check messages at time $t = 4, 8, 12, \dots$.  We take into account the temporal dynamics of user activities in our model by considering user message checking time  $\tau_i$.

\subsection{Assumptions}
	We make the following assumptions:
	\begin{enumerate}
		\item To simplify the discussion of the model, we assume each user is associated with only one device that is used to access the social network such as a smart phone, tablet, laptop or desktop. (In practice a user may have access to multiple devices, such as a desktop computer at work [owned by the employer] and a smart phone [personally owned by the user]. In this case, the user usually avoids using the employer-owned computer for personal use.) 
		
		\item  When we say \quotes{a user is infected}, we mean that the user unknowingly downloaded and executed the malware (step 2 in Section III), and his computer is subsequently infected.
		
		\item Each user has his/her own device to access the social network (i.e., no two users share the same device.). 
		
		\item We assume a single infiltrating user in the first stage of the propagation process described Section \ref{experiment}. In our future work, we will extend our model to support multiple infiltrating nodes. Note, however, that the use of many infiltrating nodes may trigger early detection of the malware in the network.
		
		\item  We assume that a new malware $M$ appears and starts propagating at time $t = 0$.
		\end{enumerate}
		
		In addition to the above assumptions, we also assume that a percentage of the OSN population, $\beta_{max}$, have AV programs installed on their systems. According to a survey conducted by Microsoft, 75\% of the respondents reported that they installed AV products on their computers \cite{Microsoft}.  We will assume $\beta_{max} = 0.75$ in our experiments. 	Note, however, that only a subset of these AV products are effective against the new malware $M$.  In other words, only a percentage of the OSN population, $\beta_{min}$, where $\beta_{min} \leq \beta_{max}$, are immune to malware $M$ at time $t = 0$. (Analogically, a flu vaccine may not be effective against all the flu strains, especially new strains.)  The value of $\beta_{min}$ depends on the novelty and sophistication of malware $M$.  The more novel and sophisticated malware $M$ is, the lower the value of $\beta_{min}$.
		
Infected users may seek clean-up solutions to disinfect their systems by themselves (independent disinfection), or with help from their friends from the social network (collaborative		 disinfection). The issues of collaborative disinfection and independent disinfection will be discussed in detail in Sections~\ref{exp3} and \ref{exp4}, respectively.
		
		\subsection{Objective of the Model}
		The objective of the model is to estimate the expected number of users $E_X(t)$ in each state $X = \{Sus, Inf, Rec, Imm\}$ at each time unit $t$. 
			\begin{equation*}
				E_X(t) = \sum_{i=1}^{i=V} P(X_i(t)=X),
			\end{equation*}
			\noindent
			where $P(X_i(t) = X)$ denotes the probability of user $i$ being in state $X$ at time $t$. For example, $E_{Inf}(t) = \sum_{i=1}^{i=V} P(X_i(t)=Inf)$ gives the expected number of infected users in the network at time $t$.
		
		To compute the probability of user $i$ being in state $X$ at time $t$, $P(X_i(t) = X)$, we need to compute the probabilities of user $i$ transitioning from one state to another, namely probabilities $\gamma_i(t)$, $\beta_i(t)$ and $\alpha_i(t)$ shown in Fig. \ref{pm}.   
		In Sections \ref{transitstoi} to \ref{transinfimm}, we discuss the computations of $\gamma_i(t)$, $\beta_i(t)$ and $\alpha_i(t)$, respectively.  In Section \ref{compmodel}, we discuss the computation of $P(X_i(t)=X)$ and $E_X(t)$, where $X = \{Sus, Inf, Rec, Imm\}$, to complete the model. To facilitate the description of the model, we summarize the mathematical notations in Table \ref{sum}.
		
\section{The Proposed Malware Propagation Model}
\label{model_detail}

In this section, we discuss the computations of the transition probabilities $\gamma_i(t)$, $\beta_i(t)$ and $\alpha_i(t)$, and the expected number of users in each  state $X$ where $X = \{Sus, Inf, Rec, Imm\}$.

\subsection{Transition from Susceptible to Infected State}
\label{stoi}
\label{transitstoi}
Let $\gamma_i(t)$ denote the transition probability of node $i$ from the susceptible state to the infected state at each time $t$. This transition probability depends on two factors:
\begin{itemize}
	\item The states of node $i$'s neighbors. The more infected neighbors $i$ has, the higher the probability $i$ will get infected. 
	\item The probability $p_i$ of user $i$ executing the malware. The higher the probability $p_i$, the higher the probability $i$ will get infected.
\end{itemize} 

\noindent Therefore, the probability $\gamma_i(t)$ of user $i$ transitioning from the susceptible to the infected state is given by the following equation:
\begin{equation}
\label{eq1}
\gamma_i(t) =1 - \prod_{j \in N_i}(1- p_iP(X_j(t-1)=Inf))
\end{equation}
where $N_i$ is the set of node i's neighbors.

Note that probability $p_i$ depends on several factors such as the probability of user $i$ viewing the malicious link on her wall (or in the private message box), the trust level between user $i$ and her friend $j$ who posted the link, the probability of following the malicious link by clicking on it, the probability of downloading the malware and the probability of executing the malware. In Appendix A of the supplementary file, we discuss the factors that affects probability $p_i$ in detail.

\begin{table*}
	\centering
	\caption{Mathematical notations}
	\label{sum}
	\begin{tabular}{| c | p{12cm} |}
		\hline
		\multicolumn{1}{|c|}{\textbf{Parameter}} & \multicolumn{1}{c|}{\textbf{Description}} \\ \hline
		V 		&  Total number of nodes (users) in the graph (network). \\ \hline
		$X_i(t)$ & State of the node $i$ at each time $t$ \small{(see Eq. (\ref{eq0}))}\\ \hline
		$p_i$		 & Probability of user $i$ following a malicious post, unknowingly downloading the malware and executing it. This probability depends on several factors as discussed in Section \ref{transitstoi}. \\ \hline
		$N_i$	& Set of node $i$'s neighbors.  \\ \hline
		$d_i$	& Degree of node $i$  \\ \hline
		$\delta_{i}$		 & Probability of user $i$ accepting clean-up solutions from their non-infectious friends (Section \ref{transinfimm}).     \\ \hline
		$q_i$	& Probability of user $i$ recovering independently without assistance from his/her friends (Section \ref{transinfimm}).  \\ \hline
		$\pi_{i}$		 & Probability of an infected user $i$ taking no action to remove malware. That is, $\pi_i=1-q_i-\delta_i$ (Section \ref{transinfimm}).    \\ \hline
		$\gamma_i(t)$	  & Probability of user $i$ transitioning from the susceptible to infected state.  This probability depends on the number of friends user $i$ has and probability $p_i$ listed above (Section \ref{transitstoi}). \\ \hline
		$\beta_i(t)$	 & Probability of user $i$ transitioning from the susceptible to immune state.  This probability depends on the effectiveness of user $i$'s AV software, or the availability of AV updates to user $i$.  [Some AV  product vendors release updates against malware $M$ sooner than others (Section \ref{sus2immune})] \\ \hline
		$\beta_{max}$ & Percentage of the social network population that has AV products installed on their computers.  [However,at the beginning of the propagation only a subset of these AV products, $\beta_{min}$, are effective against the new malware $M$ (Section \ref{sus2immune}).] \\ \hline
		$\alpha_i(t)$   & Probability of user $i$ transitioning from the infected  to recovered state. This probability depends on the number of user $i$ infected friends at time $t$ and probability $\delta_i$ listed above.  The higher these values, the higher the probability $\alpha_i$ (Section \ref{transinfimm}). \\ \hline

	\end{tabular}
\end{table*}

\subsection{Transition from Susceptible to Immune State}
\label{sus2immune}
Users can benefit from antivirus (AV) software products to protect themselves against malware attacks. Users install AV software either proactively to prevent infections, or reactively to disinfect their systems after being infected.   In the former case, an effective AV software against a malware $M$ allows a user to transition from the  susceptible to the immune state (with respect to malware $M$).  In the latter case,  up-to-date AV products enable users to disinfect themselves,  transitioning from the infected to the recovered state.   In this section, we focus on the former case (susceptible to immune state), while the latter case (infected to recovered state) is discussed in Section~\ref{transinfimm}.

If user $i$ has an AV product installed, it may or may not be effective against the new malware $M$ when $M$ first emerges in the network at time $t=0$.  If an AV product is effective against $M$, the user is considered to transition from the  susceptible to the immune state at time $t=0$.  

In practice, many malware programs use novel techniques to evade detection by AV software \cite{evas1, evas2}.  However, as a malware spreads through a network, AV manufacturers respond by providing software updates and clean-up solutions (via either updated signatures or heuristic techniques \cite{sigheu}).   Not all AV companies can provide detection and clean-up solutions at the same time though \cite{avtest}. In fact, there are cases where it takes AV providers several days to come up with disinfecting solutions. For instance, Microsoft Malware Protection Center provided detection and clean-up solutions against the Conficker malware \cite{conficker} on November 21, 2008 \cite{MMPConficker} while many other vendors such as Sophos and Trend-Micro released disinfection solutions to their users several days later, on November 26, 2008 \cite{trendconfick}.

If an AV product is not effective against the malware in the first stages of its propagation, the user can still have a chance of getting to the immune state if the AV manufacturer releases working updates and the user's AV software is updated before he/she is infected.   Because AV providers may not be able to release product updates right away,  we assume that the rate at which AV products are updated can be a function such as linearly increasing or exponentially increasing.   Therefore, we define  $\beta_i(t)$, the probability of a node $i$ transitioning from the susceptible to the immune state, as follows:

\begin{equation}
\label{eps2}
\beta_i(t) = \begin{cases}
\hat{\beta}(t) &  \text{if $0  < t \le T_{max}$} \\
\beta_{max} &  \text{if $t \ge T_{max}$}
\end{cases}
\end{equation}
where $\beta_{max}$ is the (maximum) percentage of the population that has AV products installed on their computers (assumption 6, Section~\ref{modeldes}). Only a subset of this population, represented by parameter $\beta_{min} \leq \beta_{max}$,  has AV products that are {\em effective against} $M$ at time $t=0$. As AV manufacturers gradually release updates against $M$, all AV users's products will eventually be effective against $M$.  $T_{max}$ is the time at which all AV products have been updated and able to block/remove malware $M$. 

$\hat{\beta}(t)$ denotes the percentage of users at time $t$ who have effective AV software against the new malware \textit{M}, and $\beta_{min} \le \hat{\beta}(t) \le \beta_{max}$. $\hat{\beta}(t)$ depends on the rate at which AV products are updated, and thus can be a function such as linear or exponential increase, where $t$ denotes the time unit passed.   As a linearly or exponentially increasing function, it can be represented as  $\hat{\beta}(t)=c_0 \times t$ or $\hat{\beta}(t)=c_1e^{c_2 \times t}$ respectively, where $c_0, c_1$ and $c_2$ are constants.   The value of $\hat{\beta}(t)$ increases until it reaches  $\beta_{max}$ at time $T_{max}$, when $\beta_{max}\times V$ users have effective AV software against the new malware $M$, where $V$ is the total number of users in the network. 

\begin{figure}
	\centering
	\includegraphics[width=3.0in]{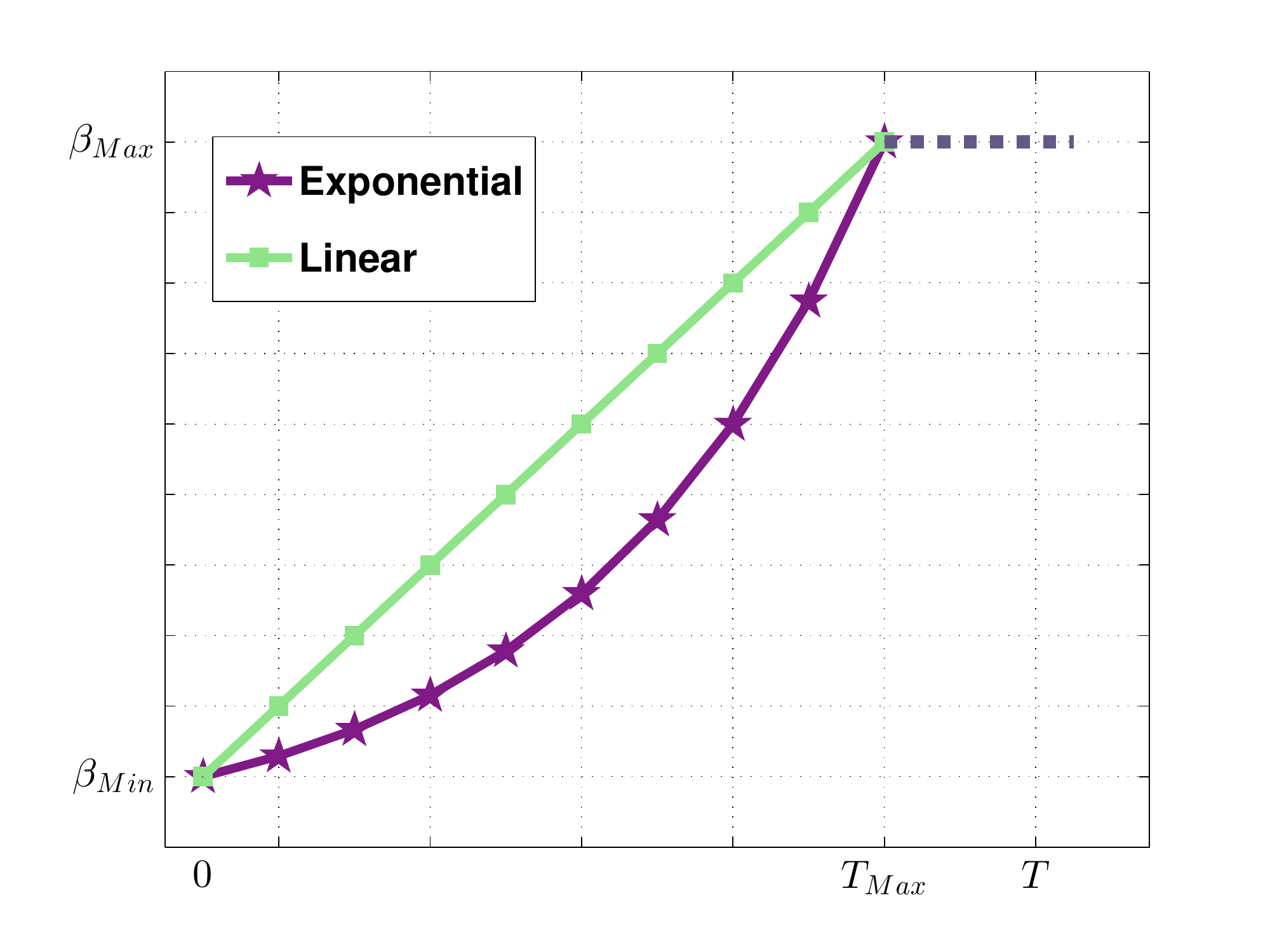}
	\caption{$\hat{\beta}(t)$ as linear and exponential functions}
	\label{dist}
\end{figure}

\subsection{Transition from Infected to Recovered State}
\label{transinfimm}
As discussed in the previous sections, a user can be infected by not having an AV product, or having an AV product that is not effective against the new malware. In the former case, the user would seek clean-up solutions in order to disinfect his OSN profile and computer.   In the latter case, the infected user's AV product would need to be updated in order to remove the new malware. 

Recent malware has presented a major obstacle to users seeking clean-up solutions or AV updates:  such a malware would block users' access to web sites of reputable AV software providers.  This obstacle has been observed in several malware attacks~\cite{koobface,kasper,extmalware,ghost,sophoslab,conficker}, including those on Facebook such as Magnet \cite{kasper}, Koobface \cite{koobface}, and extension-based malware \cite{extmalware} that targets Facebook users' browsers.   By blocking users' access to web sites of AV software providers or disabling their AV software,  attackers can maintain their control of the infected systems as long as possible to carry out malicious activities.  In Appendix B of the supplementary file, we discuss techniques malware creators use to prevent infected systems from being connected to AV provider web sites.

In this case, an infected user would have to search for clean-up solutions or AV updates via means other than direct access to or download from AV provider web sites. Following are examples of such methods:

\begin{enumerate}
	\item One method is to search for clean-up solutions from third-party web sites (which are not blocked by the malware). This method carries the risk of downloading another malware disguised as an AV product \cite{sophos,fakeAV} and is recommended only to knowledgeable users.
	
	\item A safer alternative is for the infected user to seek assistance from his OSN friends, especially those who were infected but have found effective disinfecting products and have recovered.  This practice was observed during a large-scale  attack on Facebook caused by Magnet malware \cite{kasper} that left more than 110,000 users infected in less than two days \cite{kasper}. During the attack, infected users were blocked by the malware from accessing AV provider web sites. As a result, they posted messages on social media, seeking help from their OSN friends to remove the malware from their systems.  Clean-up solutions suggested by trusted friends have been tested by them and therefore usually safe and effective.  We term the practice of seeking clean-up solutions from OSN friends \textit{collaborative disinfection}. This method is the safest and easiest for less sophisticated computer users.
	
	\item Another safe method is to access AV provider web sites directly to download AV updates using a second, clean computer and then transfer the AV update to the infected computer. (However, not all users have access to a second, clean computer.) We call this practice and method 1 discussed above \textit{independent disinfection} (as opposed to collaborative disinfection with help from friends). 
\end{enumerate}
We consider both independent disinfection (method 1 and 3) and collaborative disinfection (method 2) when calculating the probability $\alpha_i(t)$ of user $i$ transitioning from the infected  to the recovered state. 

Let $\delta_i$ and $d_i$ represent the probability of user $i$ accepting clean-up solutions from his/her friends and the degree of node $i$, respectively. Furthermore, let $q_i$ and $N_i$ denote the probability of user $i$ recovering without help from his/her friends (independent disinfection), and the set of node $i$'s neighbors respectively.  The probability $\alpha_i(t)$ of user $i$ transitioning from the infected  to the recovered state is given by the following equation:

\begin{equation}
\label{eq2}
\alpha_i(t) =  q_i + \frac{\delta_{i}}{d_i}\sum_{j \in N_i}(1-P(X_j(t-1) = Inf))
\end{equation}

Eq. (\ref{eq2}) takes into account both independent disinfection (parameter $q_i$) and collaborative disinfection (the second term of the equation). In the case of collaborative disinfection, the higher the probability that user $i$ accepts clean-up solutions from his/her friends, the higher his/her chance of recovering.  The more non-infected friends user $i$ has, the higher his/her chance of getting clean-up solutions from them to recover.  (It is logical to assume that infected friends cannot help.  If they had knowledge of clean-up solutions, they would already have disinfected their systems and moved to the recovered sate.)

So far we have discussed the transition probabilities based on the model illustrated in Fig. \ref{pm}.   In the next subsection, we provide the complete model characterizing the propagation of a Trojan malware in an OSN.

\subsection{The Complete Model}
\label{compmodel}

A social network is represented by an undirected graph $G = (\mathcal{V, E})$ where each vertex $v \in \mathcal{V}$ represents a user and each edge $e \in \mathcal{E}$ denotes the mutual relationship between the two users represented by the two end vertices.

For each user $i$, the network graph provides the following information: node degree $d_i$ and the set of user $i$'s neighbours $N_i$.
Each user $i$ is also associated with a set of parameters in the form of a vector  $\{p_i, \delta_i, q_i \}$.   Brief descriptions of these parameters can be found in Table \ref{sum}. At time t = 0 (when the malware $M$ first appears at the infiltrating node), the probability of a user $i$ being in a state $X$ is as follows, where $X = \{Sus, Inf, Rec, Imm\}$. For the infiltrating node, denoted by $k$, $P(X_k(0)=Inf) = 1$ because $k$ is considered the first node in the network infected with malware $M$.  $P(X_k(0)=X) = 0$ for every other state $X$, namely, susceptible, immune and recovered.



We assume that AV providers release updates against malware $M$ over time according to a function $\hat{\beta}(t)$ (see Section \ref{sus2immune}). 

Given the above initializations, Equations (\ref{eq1}) to (\ref{eq2}) and the transition model illustrated in Fig. \ref{pm},  we calculate the probability of a user $i$ being in the susceptible, immune, infected, and recovered state as follows, with brief explanations given in Table \ref{statedef}.
\\
\noindent
\begin{equation}
\label{ss}
\begin{split}
P(X_i(t+1)=Sus) = (1-\beta_i(t)) P(X_i(t)=Sus) +  &  \\
P(visit_i(t)=true) \ \gamma_i(t) \ P(X_i(t)=Sus)  & 
\end{split}
\end{equation}
\noindent
\begin{equation}
\label{is}
\begin{split}
P(X_i(t+1)=Inf) = P(visit_i(t)=true)\gamma_i(t) \times & \\
P(X_i(t)=Sus)  + (1- P(visit_i(t)=true) \times & \\
\alpha_i(t))P(X_i(t)=Inf)
\end{split}
\end{equation}
\noindent
\begin{equation}
\label{clbd}
\begin{split}
P(X_i(t+1)=Rec) = P(visit_i(t)=true) \alpha_i(t) \times & \\
P(X_i(t)=Inf) + P(X_i(t)=Rec) & 
\end{split}
\end{equation}
\noindent
\begin{equation}
\label{aved}
\begin{split}
P(X_i(t+1)=Imm) = \beta_i(t)  P(X_i(t)=Sus) + &  \\
P(X_i(t)=Imm) & 
\end{split}
\end{equation}

\begin{table*}
	\caption{Brief explanations of Equations (\ref{ss}) to (\ref{aved})}
	\centering
	\label{statedef}
	\begin{tabular}{| m{2cm} | m{14cm} |}
		\hline
		\multicolumn{1}{|c|}{\textbf{Equation}} & \multicolumn{1}{c|}{\textbf{Explanations}} \\ \hline
		(\ref{ss})	& A susceptible user remains in the susceptible state with probability $1-\beta_i(t)$.  That is, the user has not been infected yet (as determined by probability $\gamma_i(t)$), or become immune yet, i.e., his/her AV program has not yet been updated to block the new malware $M$ (as determined by function $\beta_i(t)$). Upon visiting the social network, the user  may become infected with probability $\gamma_i(t)$. 	\\ \hline
		(\ref{is})	& Upon  visiting the social network, a susceptible user will become infected with probability $\gamma_i(t)$ as shown in Fig. \ref{pm}.  Also, an infected user can recover via independent or collaborative disinfection with probability $\alpha_i(t)$ while visiting the social network; otherwise, the user stays in the infected state with probability $1 - \alpha_i(t)$.	\\ \hline
		(\ref{clbd})	&  Upon  visiting the social network, an infected user can recover with probability $\alpha_i(t)$ via independent or collaborative disinfection.  A recovered user will stay in the recovered state for the rest of the time with probability 1 due to effective AV solutions against malware $M$ obtained during the disinfection stage.  	\\ \hline
		(\ref{aved})	& An immune user at time $t=0$ will stay immune throughout the course of the attack with probability 1.  In addition to these users, a subset of susceptible users will become immune over time thanks to their AV products being updated gradually by AV providers, as determined by function $\beta_i(t)$).  	\\ \hline
	\end{tabular}
\end{table*}

Equations (\ref{ss}) to (\ref{aved}) are calculated using Equations (\ref{eq1}), (\ref{eps2}) and (\ref{eq2}) derived earlier and summarized below:

\begin{equation*}
\label{gr1}
\gamma_i(t) = 1 - \prod_{j \in N_i}(1- p_i P(X_j(t-1)=Inf))
\end{equation*}

\begin{equation*}
\label{br1}
\beta_i(t) = \begin{cases}
\hat{\beta}(t) &  \text{if $0  \le t < T_{max}$} \\
\beta_{max} &  \text{if $t \ge T_{max}$}
\end{cases}
\end{equation*}

\begin{equation*}
\label{ar1}
\alpha_i(t) =   q_i + \frac{\delta_{i}}{d_i}\sum_{j \in N_i}(1-P(X_j(t-1) = Inf))
\end{equation*}

Using Equations (\ref{ss}) to (\ref{aved}), we calculate the expected number of users in a each state $X = \{Sus, Inf, Rec, Imm\}$ at each time $t$ as follows: 

\begin{equation}
\label{calcexpect}
E_X(t) = \sum_{i=1}^{i=V} P(X_i(t)=X)
\end{equation}


The computational complexity for computing $E_X(t)$ is $O(E)$, where $E$ is the number of edges in the network graph.  A detailed discussion of the complexity is given in Appendix C of the supplementary file.

\section{Model Validation}
\label{modelperf}

In Sections \ref{modelperf} and \ref{exp}, we evaluate the accuracy of the model for estimating the propagation speed of a Trojan malware in an OSN.  Due to the lack of real data sets for evaluating analytic models~\cite{wen}, authors of all existing works in the literature have used simulations to validate their analytical models \cite{chen, wen, email, p2p}.   We use the same approach to validate our proposed model in this article.

The simulation program was implemented using MATLAB and based on discrete-event simulation.  The propagation process was simulated as described in Section \ref{experiment}.  We used the Facebook network subgraph described in Section \ref{characs} to run the simulations and to compute numerical results based on the proposed model.   We then compare simulation results with numerical results obtained from the model.
Each data point in the graphs was averaged over 100 runs, each of which started with a different infiltrating node (user) selected randomly.

\subsection{Metrics}
\label{metrics}
To compare numerical results obtained from the analytical model with simulation results, we use the following metrics: number of infected, susceptible and protected users, respectively.  The number of {\em protected users} is the sum of the numbers of immune and recovered users.  Since both immune and recovered users are eventually protected from the new malware \textit{M}, we combine them into one group in the graphs to make the graphs more readable.

To obtain numerical results from the analytical model, we calculated the {\em expected number of users} $E_X(t)$ in each state as follows: $E_X(t) = \sum_{i=1}^{i=V}P(X_i(t)=X)$,  where $X$ denotes the state of a user $i$ and $X = \{Sus, Inf, Rec, Imm\}$. For example $\sum_{i=1}^{i=V}P(X_i(t)=Inf)$ gives the expected number of infected users in the network at time $t$. 

We compare the expected number of infected (susceptible, protected) users computed from the model with the number of infected (susceptible, protected) users obtained from the simulations.

We use the Pearson product-moment correlation coefficient \cite{pearson} for the comparison. The correlation coefficient $r$ ranges from ${-1}$ to ${+1}$, where a value of 1 (${-1}$) implies a positive (negative) perfect relationship between two variables $X$ and $Y$,  and a value of 0 implies no linear correlation between the variables.  A positive correlation means  that if $X$ increases then $Y$ increases. A negative correlation means that if $X$ increases then $Y$ decreases.  We use Pearson correlation coefficients to determine the correlation between our analytical model results and the simulation results. We expect that an accurate model should have high positive correlations with the corresponding simulation results, i.e., $r \approx 1$.

The correlation coefficient of two variables $X$ and $Y$ is calculated as follows \cite{pearson}:
\begin{equation}
\label{pearsonformulat}
r = \frac{cov(X,Y)}{\sigma_X\sigma_Y}
\end{equation}

To measure the significance of the correlation, we calculate the \textit{p-value} of the correlation coefficient. A p-value close to zero means that the correlation is \quotes{statistically significant} (i.e., rejecting the null hypothesis) \cite{pearson}.

\subsection{Simulation Process}
\label{setparam}

We implemented a Trojan malware based on the propagation mechanism discussed in Section~\ref{experiment}.   The simulation process is summarized as follows.  In the first step of each  experiment, a node (user) in the social network graph is chosen randomly as a seed for infiltration. (In practice, the malware creator may implement several fake profiles for infiltration.)  We mark this node as {\em infected}.  The infiltrating user will post a malicious link either on her wall or directly on the wall of each of her friends.   When a susceptible user $i$ sees the malicious link, she will follow the link and execute the malicious embedded code with a probability $p_i$. The probability $p_i$ reflects the fact that some people may be more cautious and do not follow the link or they do not see the link (e.g., because it was pushed far down on a page by many more recent posts).  (If user $i$ does not click on the link, she remains in the susceptible state.  An immune or recovered user will stay in the same state until the end of an experiment.)

The malware will then post malicious link on the wall of the newly infected user for her friends to see.   The above process then continues with these friends. In our simulation, the above steps repeat until the average number of infected nodes remains less than ten for four consecutive time units; that is, further propagation of the malware will not significantly increase the number of victims.  (In real life, the malware creator may stop the propagation when the number of infected nodes reaches a certain number.) At the end of each time unit, we counted the number of infected, immune and protected users and recorded them to plot the resulting graphs.

%
In all the experiments, we assume that the Trojan malware blocks users from accessing AV software providers' web sites, as discussed in Section \ref{transinfimm}.   In this case, infected users can seek assistance from their OSN friends to find clean-up solutions (collaborative disinfection) or search for clean-up solutions themselves (independent disinfection).  We study both cases in our experiments.

\subsection{Experiment Settings}
We conducted five  sets of experiments. In the first four sets of experiments, we assume that user message checking time follows an exponential distribution with mean 40, i.e., $\tau \sim E(40) $.  This parameter comes from previous works \cite{wen, email, chen}. In the fifth set, we examine the case in which users visit the social network more often on average, i.e., $\tau \sim E(20) $.  In all experiments, we assume that the probability of a user $i$ clicking on the malicious link and subsequently executing the Trojan code is $p_i=0.5$, unless otherwise stated.

Following are the settings of the experiments:

\begin{itemize}
	\item \textbf{Experiment I:} In the first experiment, we study the impact of malware execution probability $p_i=p$ by comparing two cases of $p=0.5$ and $p=0.75$. We assume that no clean-up solution is available to users ($\delta_i=0$ and $q_i=0$).  
	
	\item \textbf{Experiment II:} In the second experiment, we study the impact of gradual AV update releases by AV product manufacturers on the Trojan  propagation by examining different $\beta_i(t)$ functions as discussed in Section~\ref{sus2immune}.   We assume there is no  clean-up solution ($\delta_i=0$ and $q_i=0$) available to observe the effects of gradual AV update release on containing the malware.
	
	\item \textbf{Experiment III:} In the third experiment, we study the effects of collaborative disinfection by varying $\delta_i$ from 0 to 1.   We assume no independent disinfection ($q_i=0$).  
	
	\item \textbf{Experiment IV:} In the fourth experiment, we study the effects of independent disinfection by comparing two cases: $q_i=0$ vs. $q_i=0.5$ .   We assume no collaborative disinfection ($\delta_i = 0$). 
	
	\item \textbf{Experiment V:} In the fifth experiment, we study the impact of user message checking time on the Trojan propagation by comparing different $\tau$ distributions. We assume that users visit the social network more often on average, i.e., $\tau \sim E(20)$.  We compare this case with the case where $\tau \sim E(40) $ as used in all the previous experiments. 
\end{itemize}  

A summary of the experiments and their parameters and results are given in Table~\ref{tabsum}.  The graphs obtained from the experiments show the number of users (infected, susceptible and protected) over time, unless otherwise stated.

\begin{center}	
	\begin{table*}
		\caption{Experiment parameters and summary of results}
		\label{tabsum}
		\scriptsize
		\centering
		\begin{tabular}{ | m{4cm} | m{1.1cm} | m{4cm}| m{5cm} |}
			\hline
			Experiment & Figure & Parameters & Summary of Results \\ \hline
			Experiment I - Malware Execution Probability (Section \ref{inimm}) & Fig. \ref{fig1}  & $\beta_i$=0, $\delta_i$=0, $q_i$=0, $p_i$=0.5, $p$=\{0.5,0.75\} & With no AV protection in place, all susceptible users eventually become infected. The higher the value of $p$, the higher the number of infected users in earlier stages of propagation.
			\\   \hline
			Experiment II - Gradual AV Update Release (Section \ref{exp2}) & Fig. \ref{fig2a} and \ref{fig2b} & $\beta_i$ linearly and exponentially increases between 0 and 0.75, $\delta_i$=0, $q_i$=0, $p_i$=0.5 & Gradual release of AV updates help some susceptible nodes to become immune. However it does not have direct impact on the nodes that are infected with a blocking Trojan malware.
			\\ \cline{2-4}
			& Fig. \ref{linvsexp} & $\beta_i$ linearly and exponentially increases between 0 and 0.75, $\delta_i$=0, $q_i$=0, $p_i$=0.5 & The linear function outperforms the corresponding exponential function in terms of containing the malware propagation. The linear function allows faster AV update release, resulting in more susceptible users becoming immune in the early stages of the propagation. 
			
			\\   \hline
			Experiment III - Collaborative Disinfection (Section \ref{exp3}) & Fig. \ref{exp3sc1} to  \ref{exp3comp} & $\beta_{150} = 0.005t$, $\delta_i$=\{0.2,0.4\}, $q_i$=0, $p_i$=0.5 & Collaborative disinfection helps infected nodes to recover, resulting in fewer infected nodes in the network. The higher the probability $\delta_i$, the lower the number of infections.
			\\  \hline
		
			Experiment IV - Independent Disinfection (Section \ref{exp4}) & Fig. \ref{fig4} & $\beta_{150} = 0.005t$, $\delta_i$=0, $q_i$=\{0.2,0.4\}, $p_i$=0.5   &  Independent disinfection results in lower numbers of infected users. The higher the value of $q_i$, the lower the number of infections \\ \hline
			Experiment V - Frequency of Visit (Section \ref{exp5}) & Fig. \ref{fig5} & $\tau_i\!\!\sim\!E(\lambda)$, $\lambda=20$ $\delta_i$=\{0., 0.2\}, $q_i=0$, $p_i$=0.5, $\beta_i$=\{0, 0.005t\}   &  The higher the visiting frequency, the higher the number of infections \\ \hline
		\end{tabular}
	\end{table*}
\end{center}

\section{Experimental Results}
\label{exp}
In this section, we discuss the results obtained from the five experiments described above.

\subsection{Experiment I: Malware Execution Probability}
\label{inimm}
To study the effects of probability $p_i$ on the propagation of $M$, we consider two cases of $p=0.5$ and $p=0.75$. 
We assume there is no available clean-up solution ($\delta_i=0$ and $q_i=0$) in both cases.

\subsubsection*{Set I: $p=0.5$}

Figure \ref{exp1sc1} show the results obtained from the analytical model and the simulation for this experiment. As can be seen, the number of susceptible users decreases while the number of infected users increases over time due to the lack of effective AV protection. The number of infected nodes rises until reaching its maximum, which is the initial number of susceptible nodes.  The number of protected users stays at zero during the course of this experiment due to lack of AV protection.

\subsubsection*{Set II: $p=0.75$}
\begin{figure}
	\centering
	\subfigure[Experiment I - $p$=0.5  ]
	{
		\includegraphics[width=2.5in]{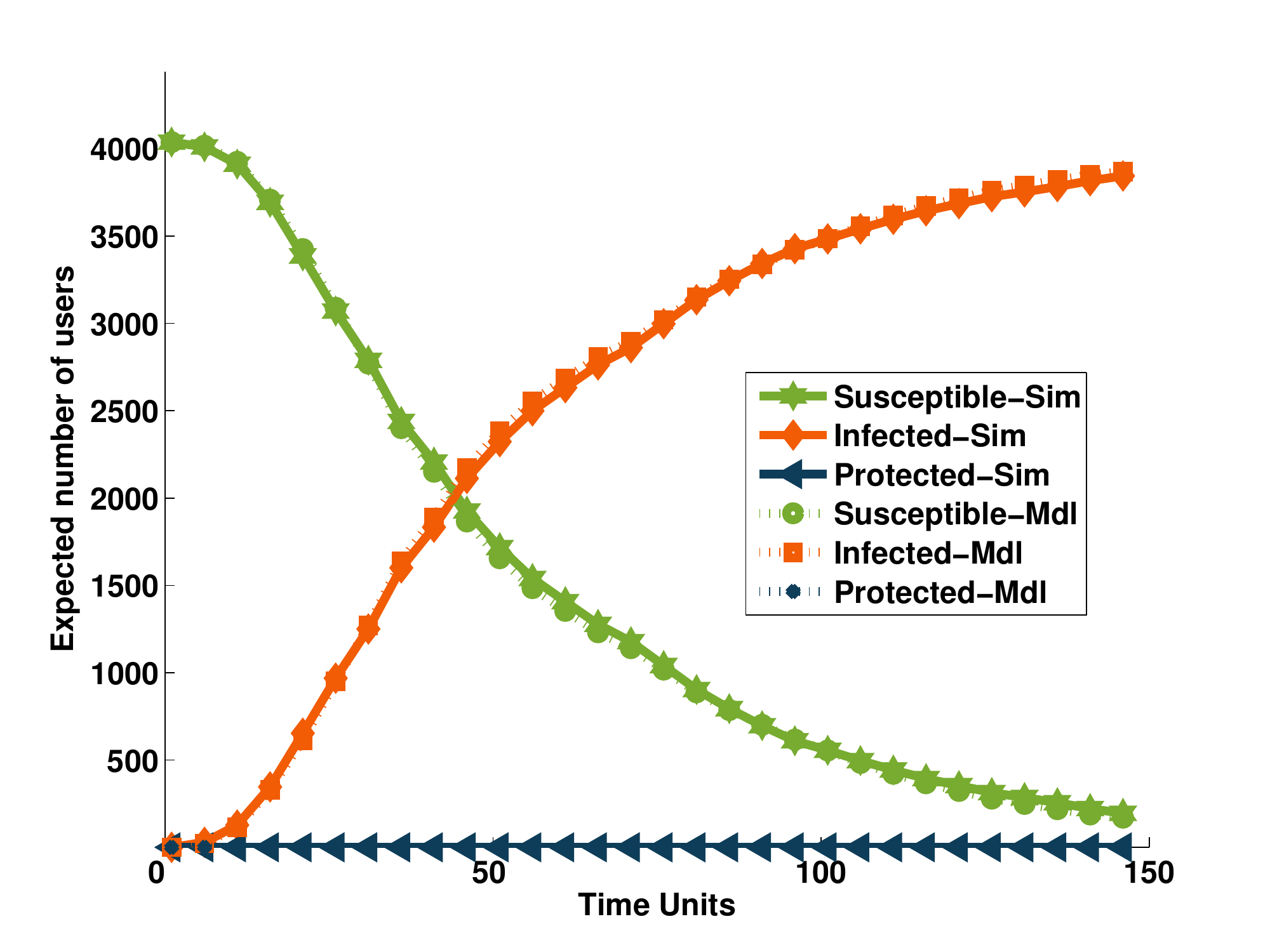}
		\label{exp1sc1}
	} 
	\subfigure[Experiment II - $p$=0.75 ]
	{
		\includegraphics[width=2.5in]{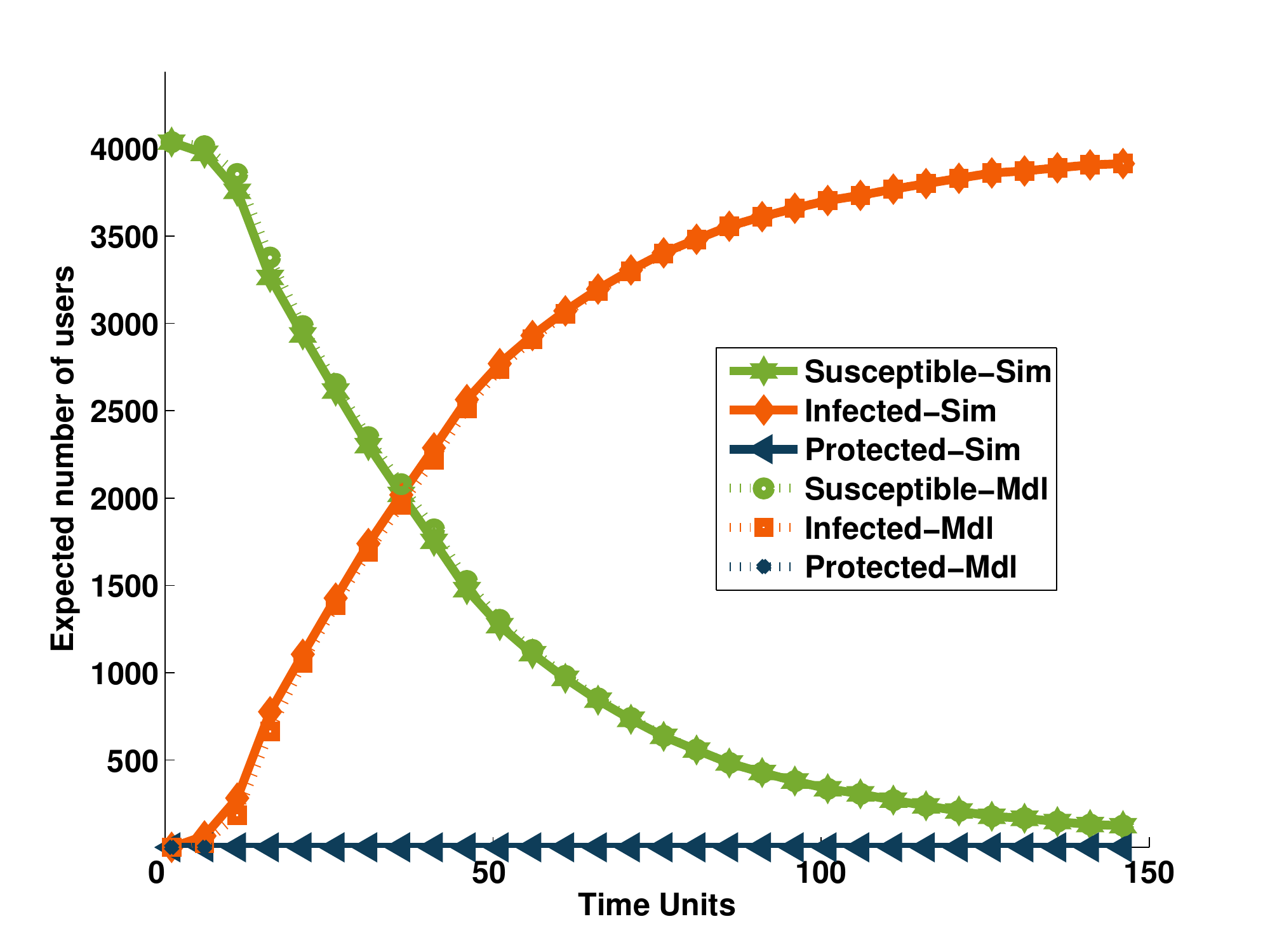}
		\label{exp1sc2}
	}
	\caption{Experiment I: Impact of malware execution probability on malware propagation - $\beta_i$=0, $\delta_i$=0, $q_i$=0}
	\label{fig1}
\end{figure}

Figure \ref{exp1sc2} shows a similar trend of users' transitions from the susceptible to the infected state.  However, since the malware execution probability is increased from $p=0.5$ to $p=0.75$, the rate of users becoming infected is higher than the previous Set. For instance, in the 50th and 100th time units, the number of infections are 2,073 and 3,306 respectively when $p=0.5$, while the these values are equal to 2,446 and 3,591 when $p=0.75$. That is, a maximum of 10\% more infection in the network at earlier stage of propagation. 


The results in Fig. \ref{exp1sc1} and \ref{exp1sc2} show that the model closely matches the simulation results. For instance, in Fig. \ref{exp1sc2}, the average error between the predicted values and the simulation results are less than 1\% for the susceptible and infected user curves, respectively. The largest discrepancy is 2\%, which occurs at the 30th time unit.  The Pearson correlation coefficient also shows a close positive correlation between the model and the simulation with $r \approx 0.99$ and $p-value \approx 0$ for  both groups of users, susceptible and infected.

In summary, without effective AV products or clean-up solutions, all susceptible users will eventually become infected.  Obviously, the higher the probability of $p_i$, the higher the number of infected users in earlier stages of propagation. 

\subsection{Experiment II: Gradual AV Update Release}
\label{exp2}
In this experiment, we study the impact of gradual AV update release on the Trojan  propagation by examining different $\beta_i(t)$ functions as discussed in Section~\ref{sus2immune}.  In particular, we consider two functions: linear and exponential.  For each function, we also examine how the rate at which AV updates are released affects  the numbers of susceptible, infected and protected users.

We assume no available clean-up solution ($\delta_i=0$ and $q_i=0$) to observe the effects of gradual AV update release on containing the malware.

\subsubsection*{Linear Functions}
Function $\beta_i$ increases linearly from 0 to $\beta_{max}$, and $\hat{\beta}(T_{max}) = \beta_{max}$.  We set $\beta_{max} = 0.75$, according to a survey conducted by Microsoft \cite{microreport} in which 75\% of the respondents said that they had AV programs installed on their systems. The value of $T_{max}$ indicates the rate at which AV updates are released.  The lower the value of $T_{max}$, the faster AV updates are released.  We consider three $T_{max}$ values: 150, 100 and 25 time units, resulting in the following three linear functions, respectively: $\hat{\beta}_{150}(t)= 0.005t$, $\hat{\beta}_{100}(t)= 0.0076t$ and  $\hat{\beta}_{25}(t)= 0.031t$.

The numerical and simulation results based on these three functions are given in Fig. \ref{exp2sc1}, \ref{exp2sc1na} and \ref{exp2sc1nb}, respectively.  

In all three cases, we observe the following:
\begin{enumerate}
	\item The number of protected nodes increases over time thanks to AV updates released gradually.
	
	\item The number of infected nodes also increases over time because we assume no clean-up solution is available.  (With clean-up solutions available to users, the number of infections will eventually go down.  We will study this case in the next experiment.)
	
	\item The number of susceptible nodes decreases over time.  They move to either the infected group (due to the malware) or the protected group (due to AV updates released gradually by AV providers).
\end{enumerate}

We note that AV updates enable a large number of users to move from the susceptible group to the protected group, resulting in less infections when compared to the case where there are  no AV updates available (compared to Fig. \ref{exp1sc1}).  For instance, the analytical result in Fig. \ref{exp2sc1} shows that there are 572 infected users at the end of the experiment while the total number of infected users in Fig. \ref{exp1sc1} is 4,039 users. That means about 85\% less infected users compared to the case where there are no AV updates, demonstrating the importance of having AV protection.

Faster AV update release limits the spread of the malware and protect more users from becoming infected.  Going from Fig. \ref{exp2sc1} to Fig.  \ref{exp2sc1na} and \ref{exp2sc1nb}, we see that the number of infected users decreases significantly.  For instance, the analytical result shows that at the 50th step, there are 206 and 65 infected users when $T_{max} = 100$ and $T_{max} = 25$, respectively.  This number is 576 infected users in Fig. \ref{exp2sc1} when $T_{max} = 150$, the slowest update rate in our experiment.   

At the same time, we observe an increase in the number of protected users as $T_{max}$ decreases from 150 to 25 (i.e., AV updates are released at faster rates).  To further illustrate this point, we consolidated the curves representing the numbers of protected users from Fig. \ref{exp2sc1}, \ref{exp2sc1na} and \ref{exp2sc1nb} and placed them in Fig. \ref{exp2sc1cons}.    We can see from the new graph that the faster AV updates are released, the higher the number of users become protected.  For instance, at the 50th step, there are 3,974, 3,833 and 3,465 protected users in the network for $T_{max} = 25$, 100 and 150, respectively.

\subsubsection*{Exponential Functions}
Similarly to the previous case, we assume $\beta_{max} = 0.75$ and consider three $T_{max}$ values: 150, 100 and 25, resulting in the following three exponential functions, respectively: $\hat{\beta}_{150}(t) \approx 0.01 \times e^{t \times 0.029}$, $\hat{\beta}_{100}(t) \approx  0.01 \times e^{t \times 0.044}$ and $\hat{\beta}_{25}(t) \approx 0.01 \times e^{t \times 0.18} $.


Figures \ref{exp2sc2}, \ref{exp2sc2nd} and \ref{exp2sc2ne} show the results respectively. The results are consistent with those  obtained from the previous set with the linear functions.  That is, as more AV products are updated, more users will become protected.  For example, in Fig.~\ref{exp2sc2nd}, the number of protected users increases from zero to 2,266 at the 50th step.

AV updates result in less infections (Fig.~\ref{exp2sc2nd}) than the case where no AV updates are released (Fig. \ref{exp1sc1}).  By comparing Fig.~\ref{exp2sc2nd} and Fig. \ref{exp1sc1}, we see 1,217 infected users versus 4,039 infected users,  a difference of 70\% at the 150th time unit.

As the rate at which AV products are updated speeds up (i.e., $T_{max}$ decreases), fewer users will be infected and more users become protected.  For example, at the 50th step, the analytical results for $T_{max} = 150$ show 1,304 infected and 2,146 protected users (Fig.~\ref{exp2sc2}), while these numbers are 364 infected users and 3,675 protected users when $T_{max} = 25$ (Fig.~\ref{exp2sc2ne}).  To further illustrate this point,  the curves representing the numbers of protected users in Fig. \ref{exp2sc2}, \ref{exp2sc2nd} and \ref{exp2sc2ne} are combined into Fig. \ref{exp2sc2cons}.   The combined graph shows 3,675, 2,822 and 2,590 protected users in the social network for $T_{max} = 25$, 100 and 150, respectively.

\subsubsection*{Linear vs. Exponential Functions}
We observe that the linear functions outperform the corresponding exponential functions in terms of containing the malware propagation.  To facilitate the comparison, Fig. \ref{exp2comp} shows the numbers of protected users obtained from the linear and exponential functions for $T_{max} = 125$ (extracted from Fig.~\ref{exp1sc1}, and \ref{exp2sc2} and \ref{exp2sc2}, respectively).  Figure \ref{exp2comp} shows that the linear function allows more users to be come protected than the exponential function: 3,465 versus 2,146 users at the 50th step, or about 30\% higher.  

\begin{figure*}
	\centering
	\subfigure[$T_{max} = 150$]
	{
		\includegraphics[width=2.5in]{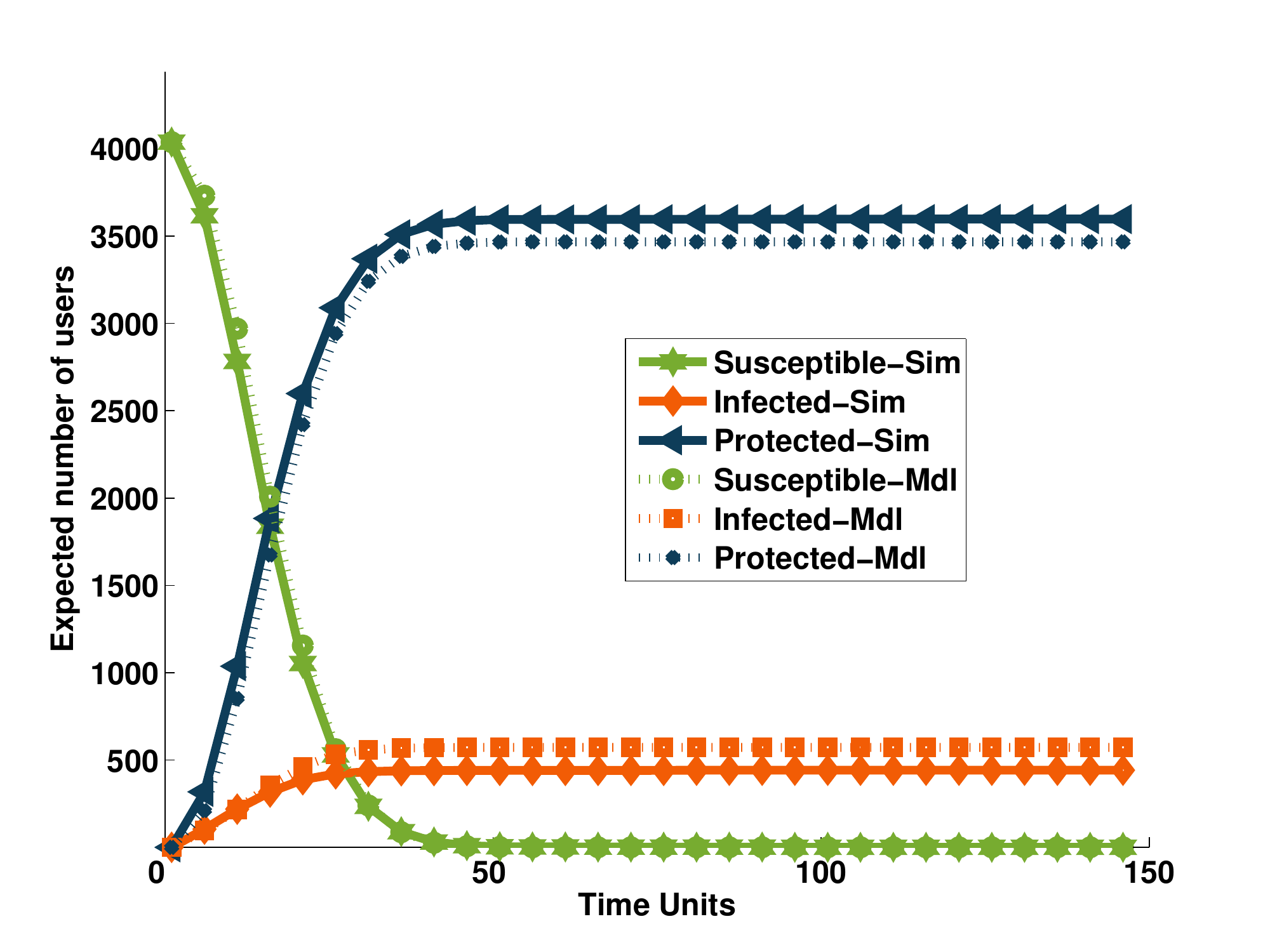}
		\label{exp2sc1}
	}
	\subfigure[$T_{max} = 100$]
	{
		\includegraphics[width=2.5in]{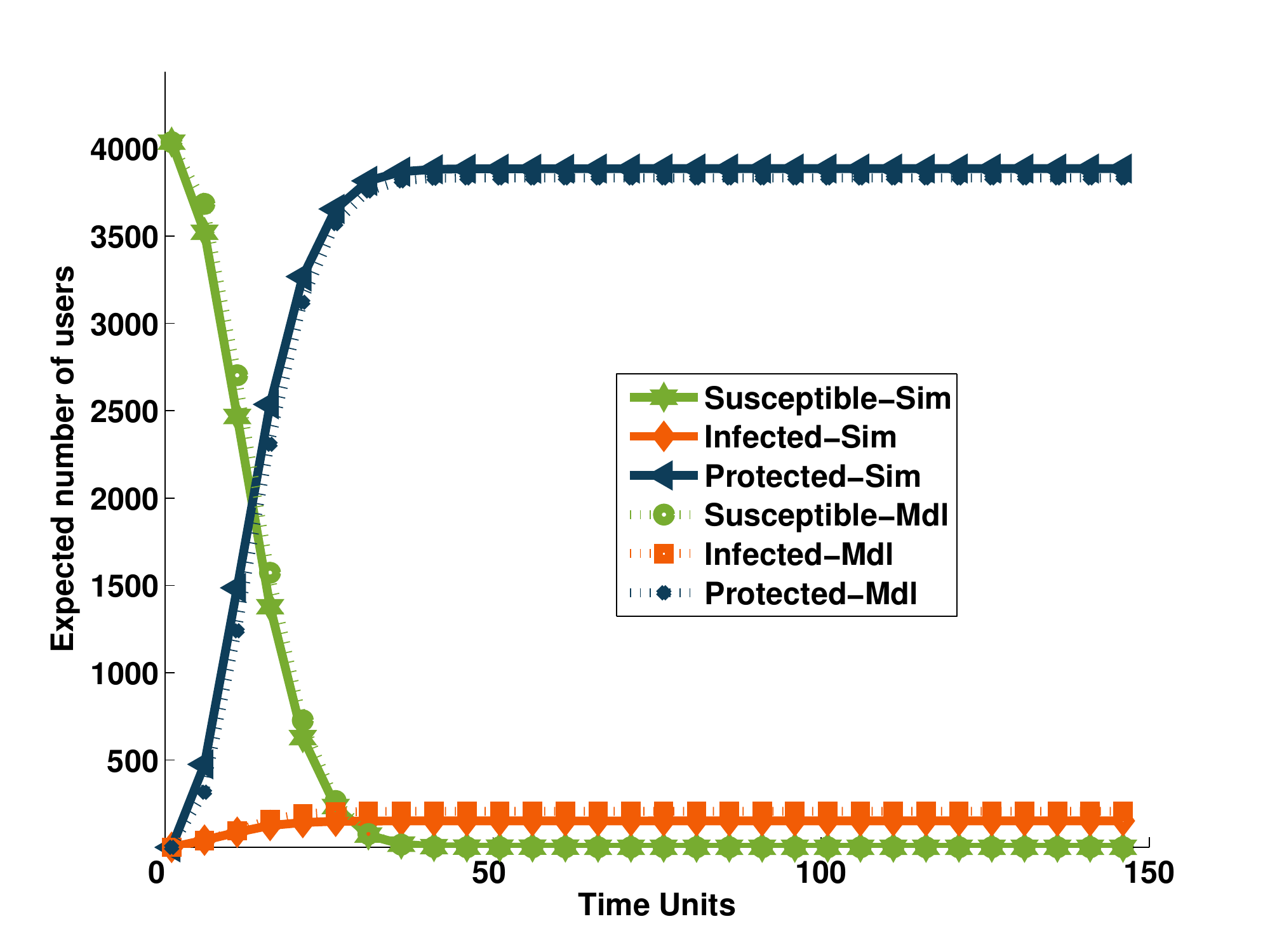}
		\label{exp2sc1na}
	} 
	\subfigure[$T_{max} = 25$]
	{
		\includegraphics[width=2.5in]{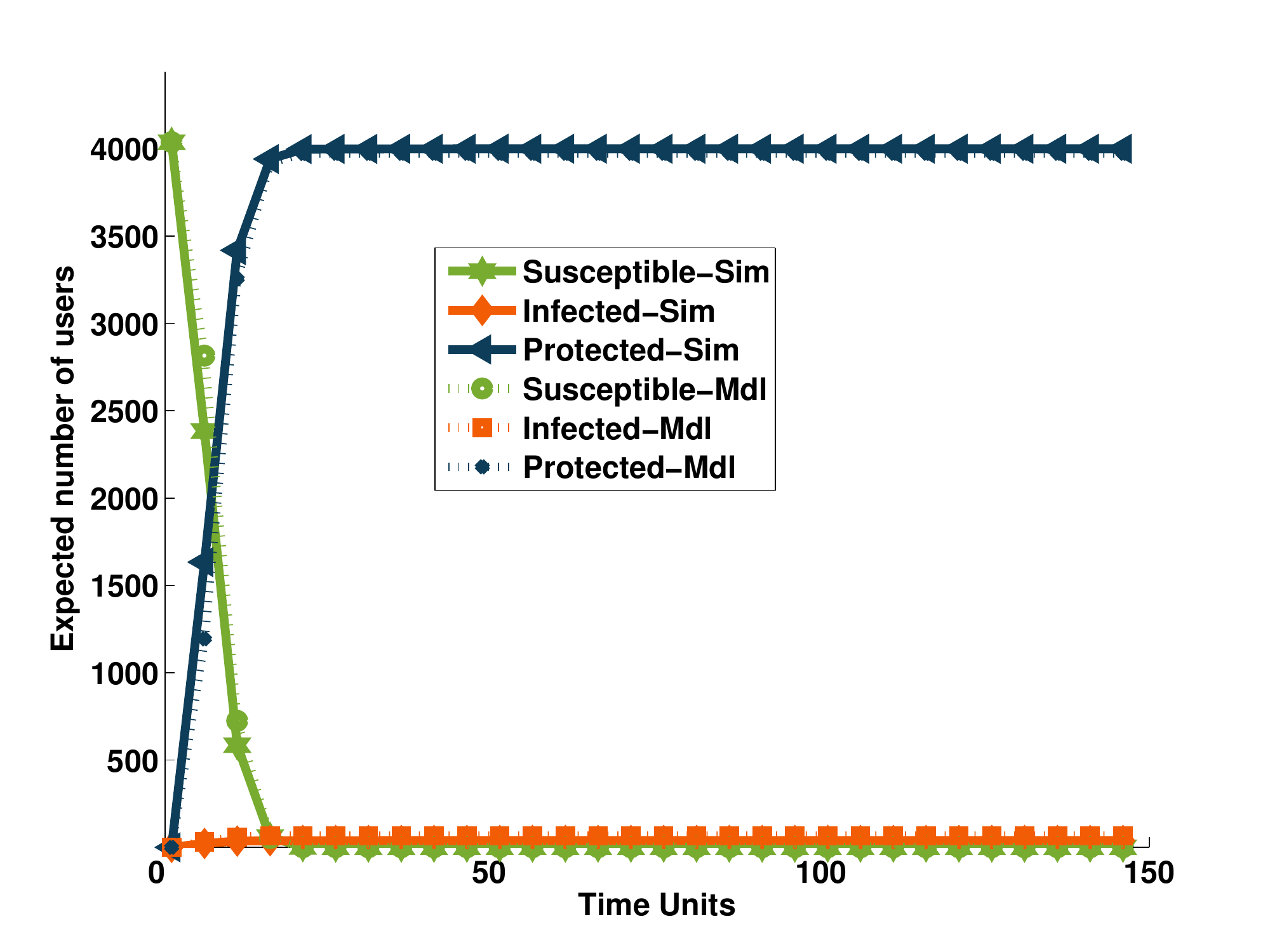}
		\label{exp2sc1nb}
	} 
	\subfigure[Consolidated graphs of number of protected users]
	{
		\includegraphics[width=2.5in]{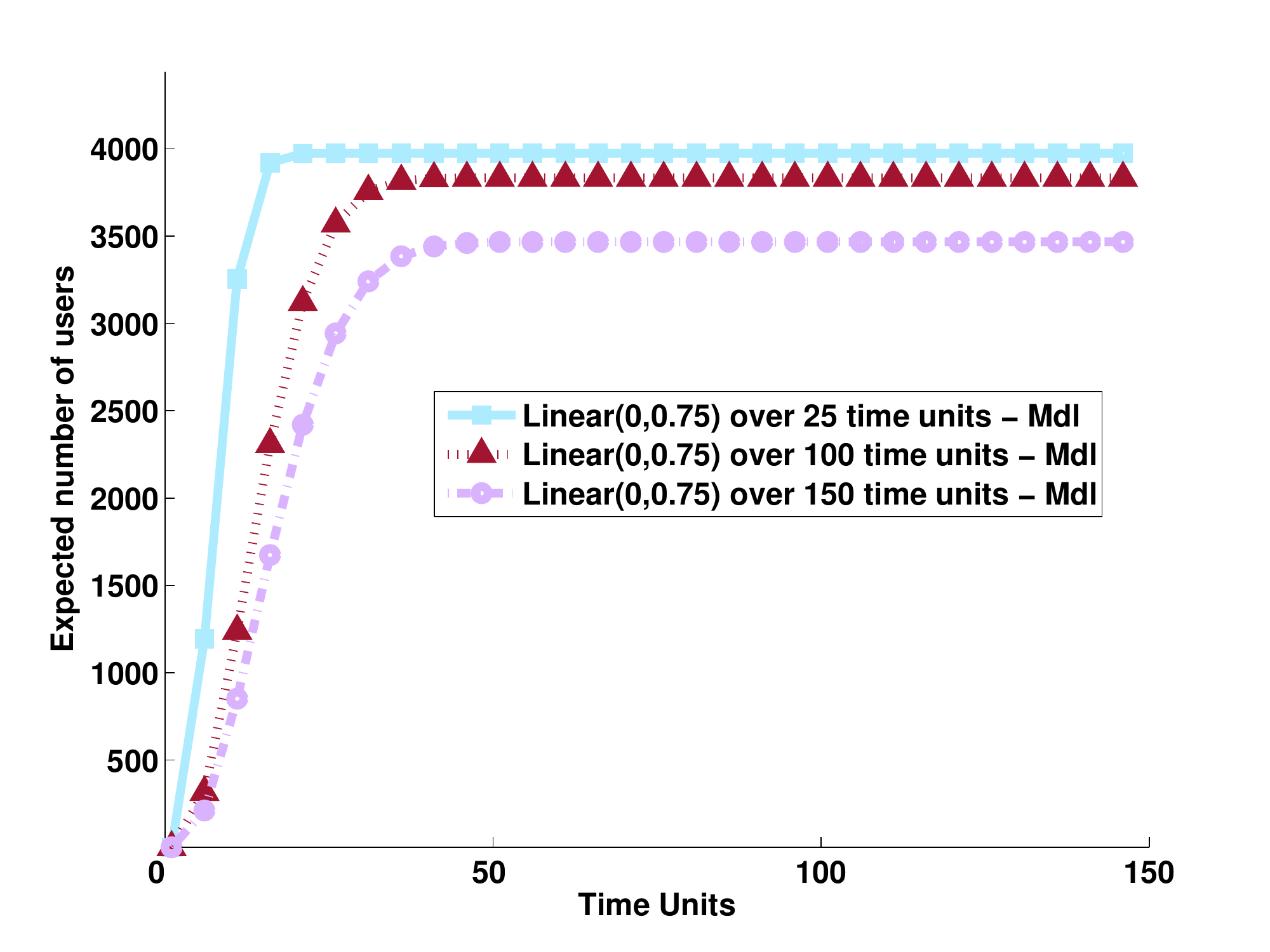}
		\label{exp2sc1cons}
	}
	\caption{Experiment II: Gradual AV update release, linear functions, $\delta_i$=0, $q_i$=0, $p_i$=0.5  }
	\label{fig2a}
\end{figure*}

\begin{figure*}
	\centering
	\subfigure[$T_{max} = 150$]
	{
		\includegraphics[width=2.5in]{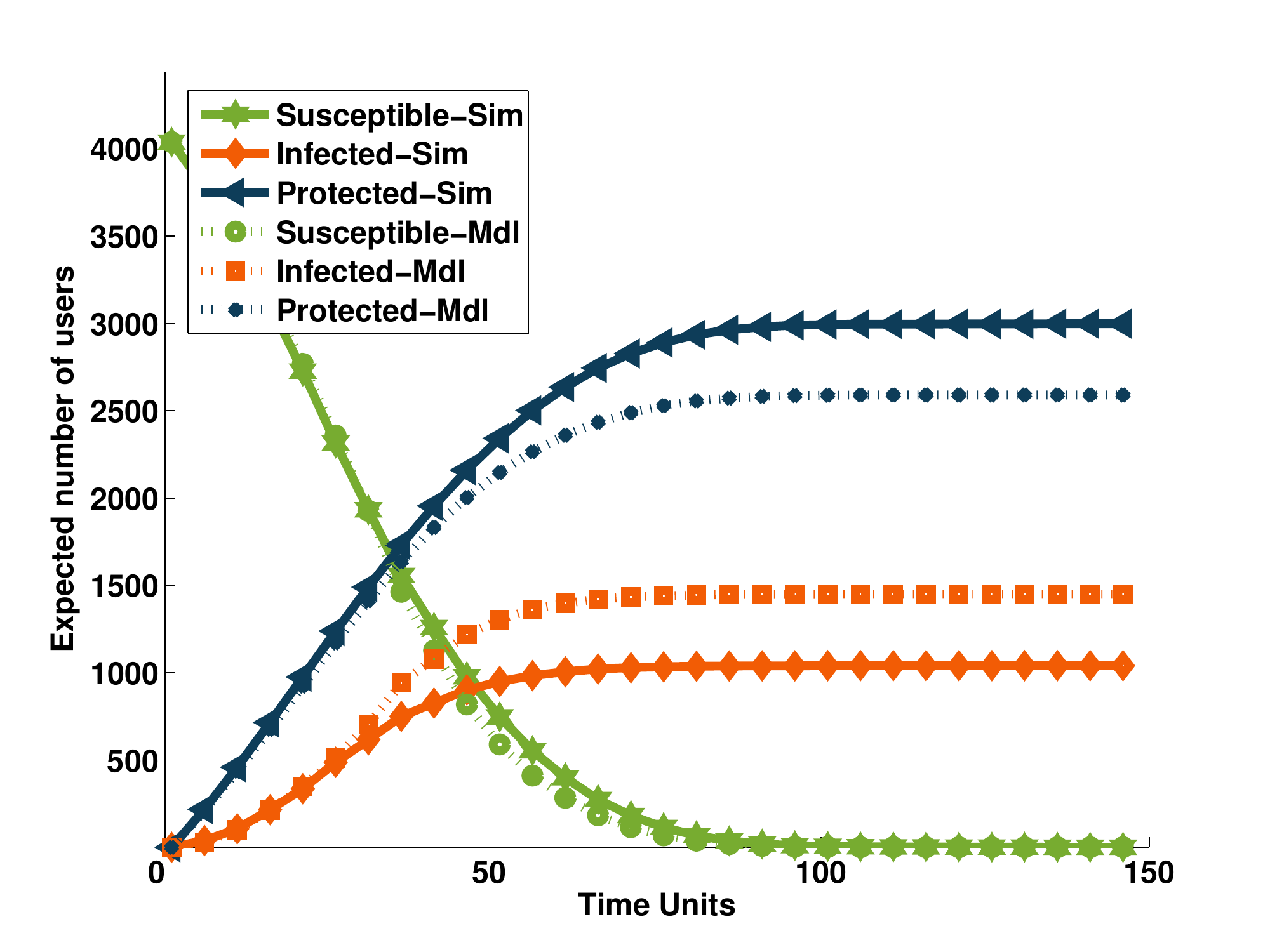}
		\label{exp2sc2}
	}
	\subfigure[$T_{max} = 100$]
	{
		\includegraphics[width=2.5in]{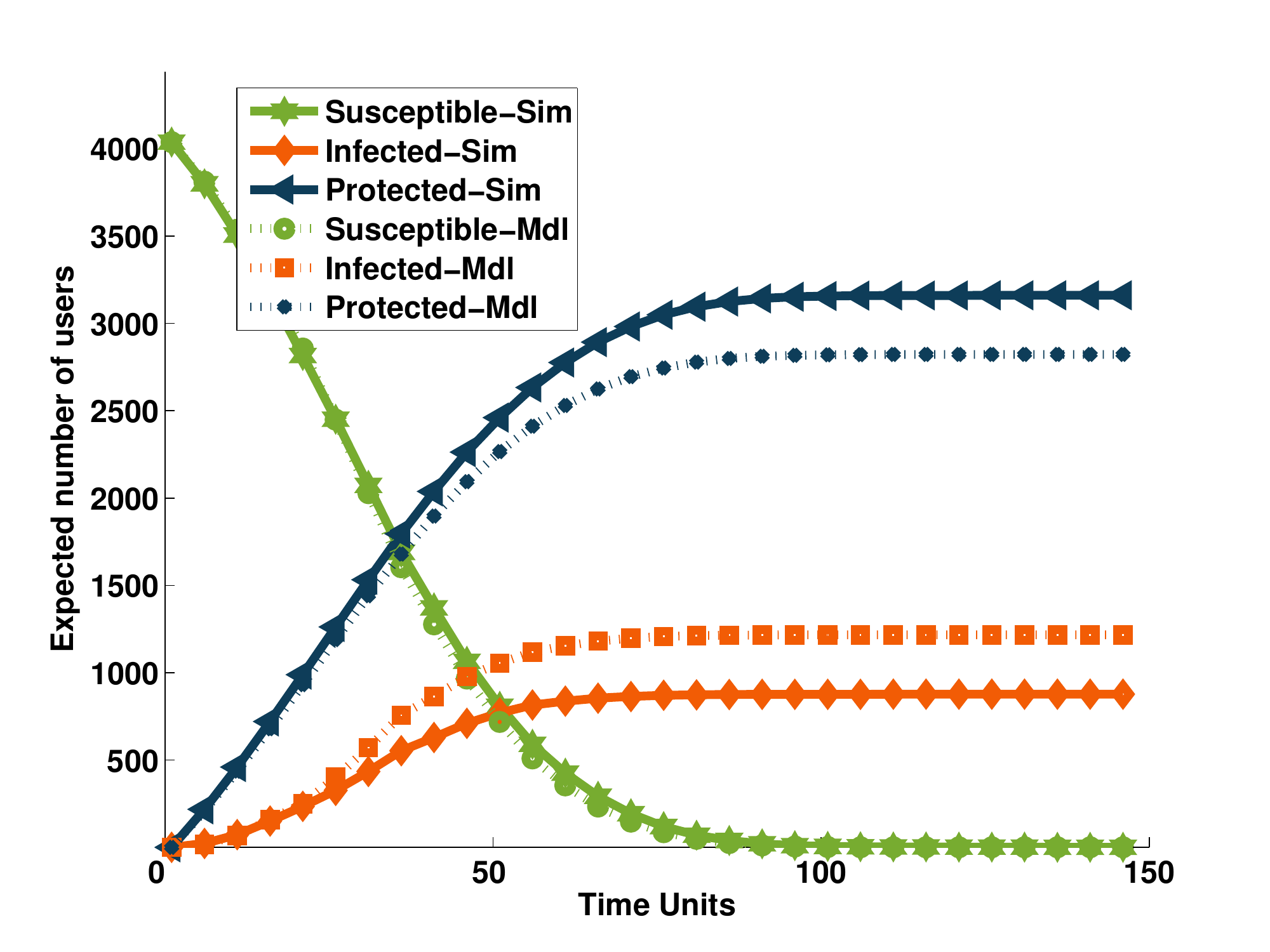}
		\label{exp2sc2nd}
	} 
	\subfigure[$T_{max} = 25$]
	{
		\includegraphics[width=2.5in]{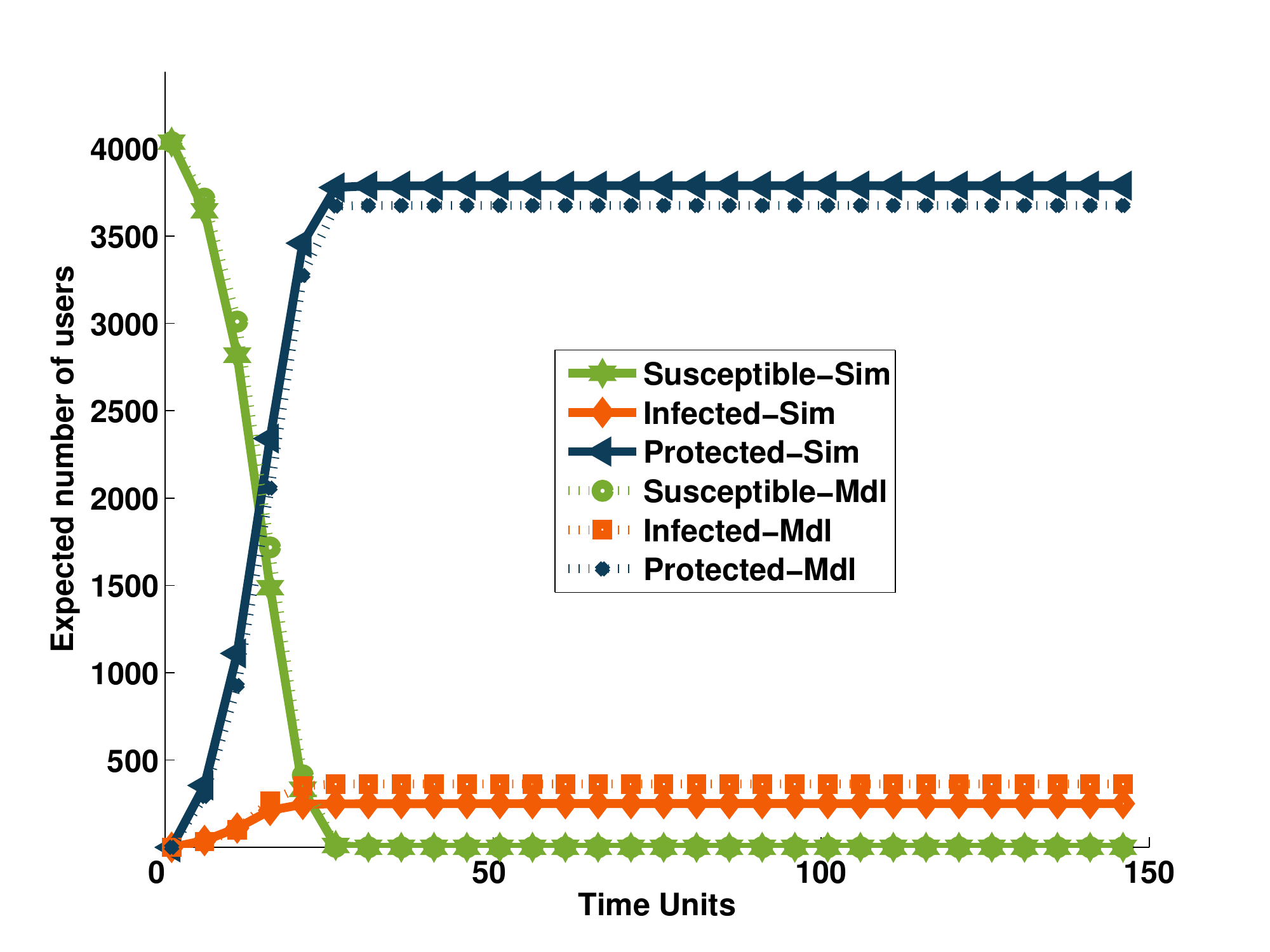}
		\label{exp2sc2ne}
	} 
	\subfigure[Cosnolidated graphs of number of protected users]
	{
		\includegraphics[width=2.5in]{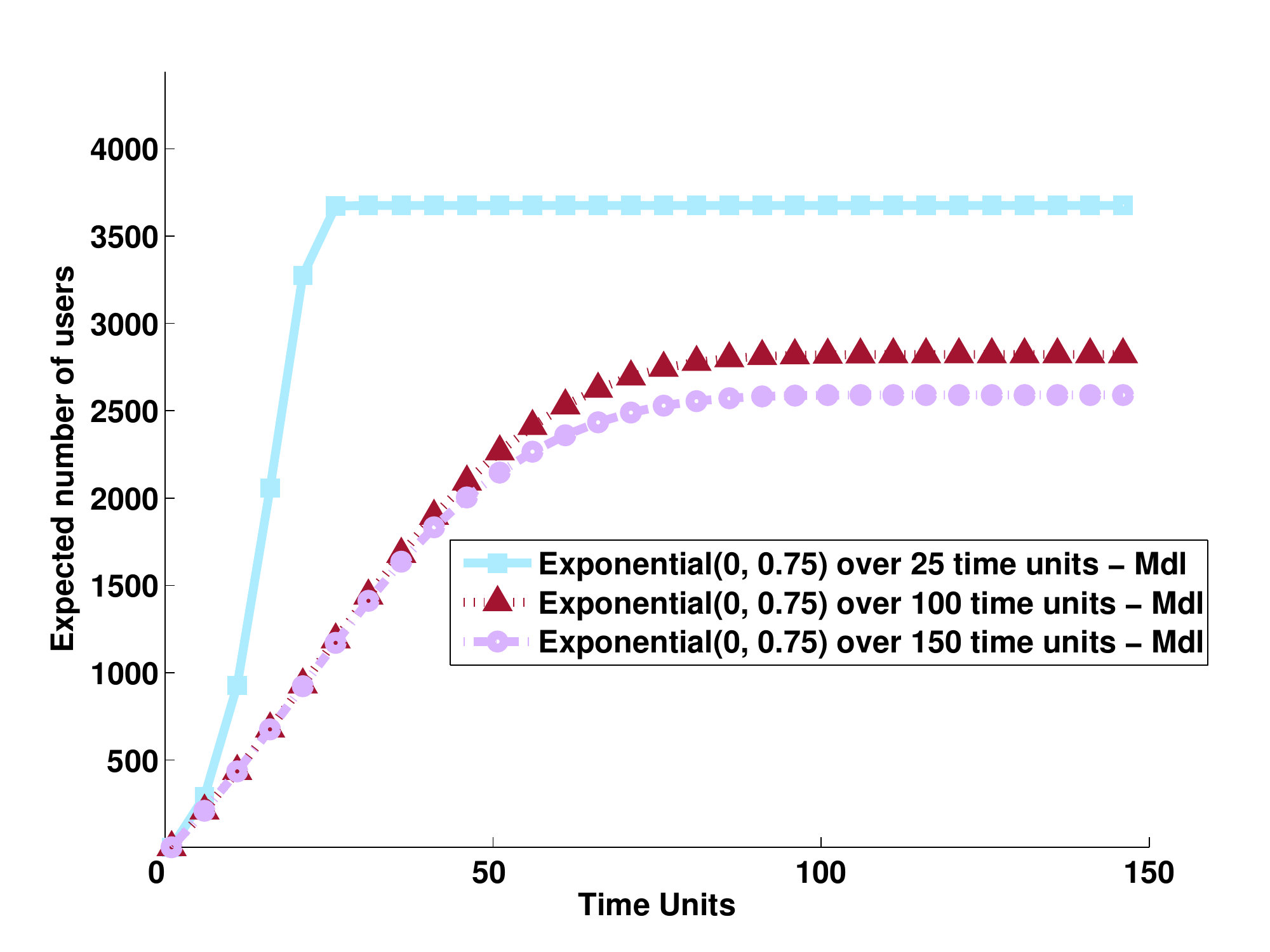}
		\label{exp2sc2cons}
	}
	\caption{Experiment II: Gradual AV update release, exponential functions, $\delta_i$=0, $q_i$=0, $p_i$=0.5  }
	\label{fig2b}
\end{figure*}

\begin{figure*}
	\centering
	\subfigure[Number of protected users]
	{
		\includegraphics[width=2.5in]{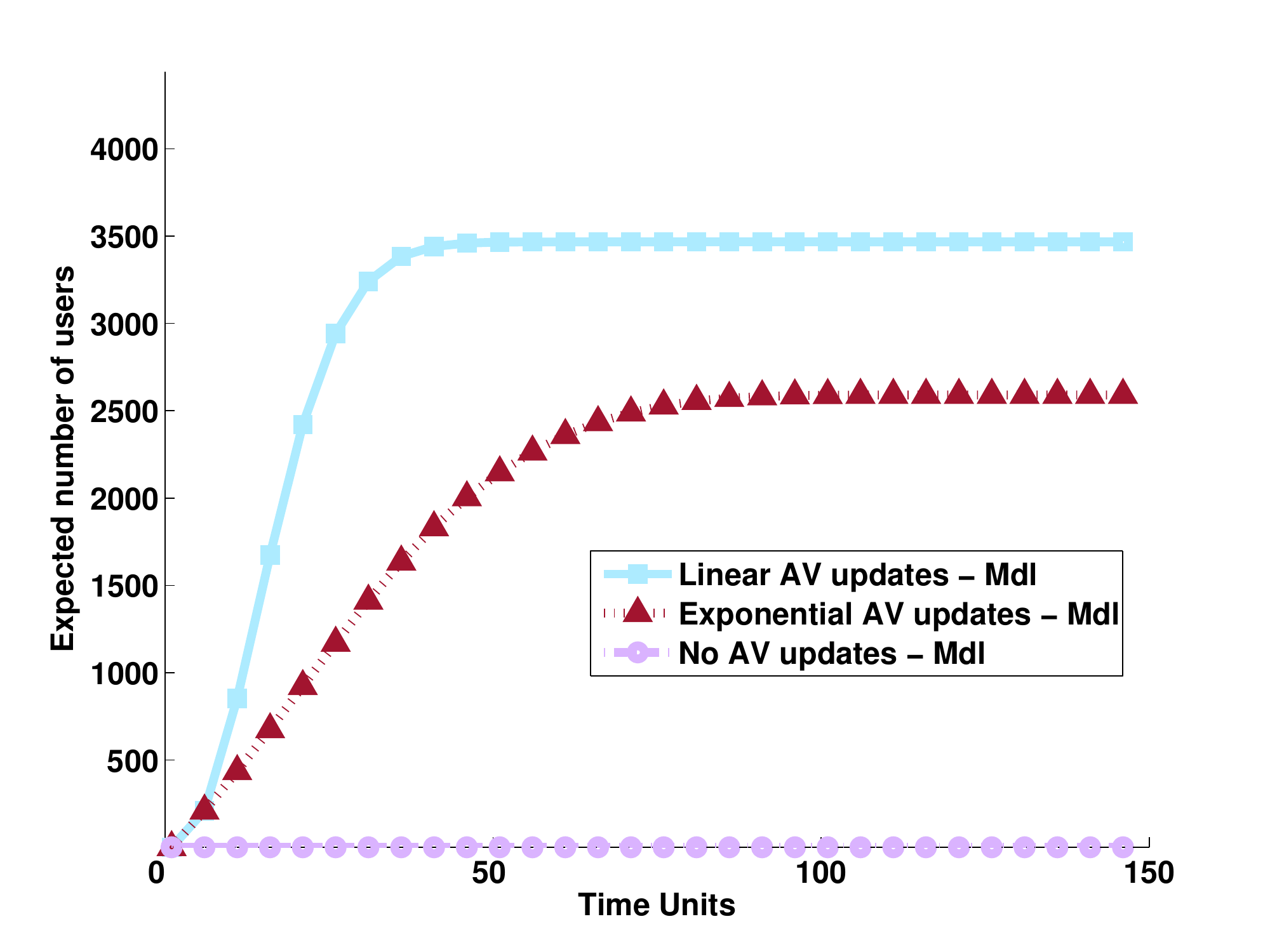}
		\label{exp2comp}
	}
	\subfigure[Gradual AV update release, different $T_{max}$ values]
	{
		\includegraphics[width=2.5in]{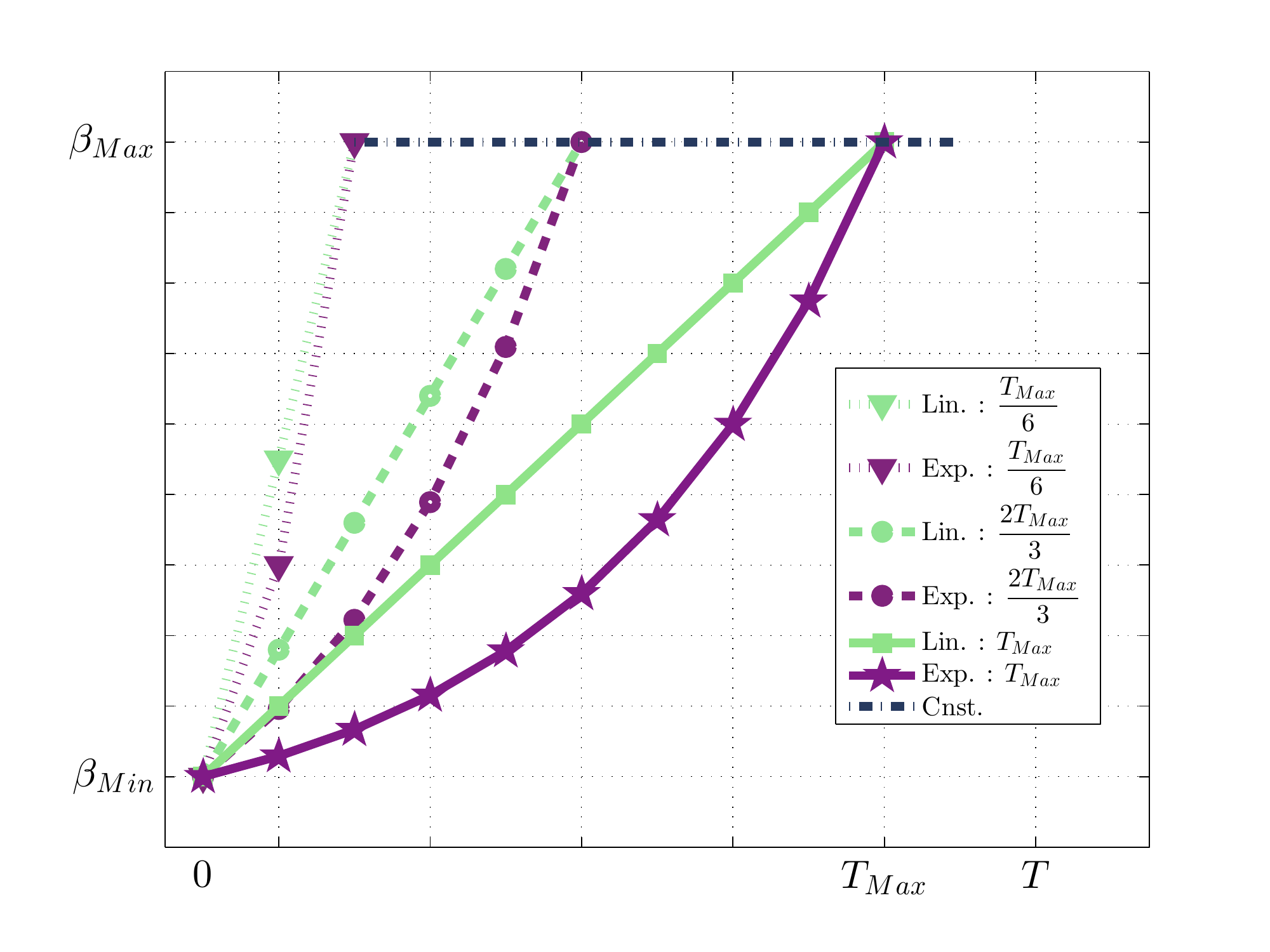}
		\label{distcomp}
	}
	\caption{Experiment II: Linear vs. exponential functions}
	\label{linvsexp}
\end{figure*}

The explanation for the above observation is that the linear function grows faster than the exponential function for $0  < t \leq T_{max}$, $T_{max} = 150$ (refer to Fig.~\ref{dist}).   The linear function allows faster AV update release, resulting in more susceptible users becoming immune in the early stages of the propagation. To visualize this comparison, we consolidated  the curves representing the number of protected users in Fig. \ref{exp1sc1}, \ref{exp2sc1} and \ref{exp2sc2} into the graph in Fig. \ref{exp2comp}.

Figure \ref{exp2comp} also includes the curve of the numbers of protected users extracted from Fig. \ref{exp1sc1} for the case of no available AV updates.   In this case, there are no immune users because there are no effective AV products against malware $M$ and no AV updates either.

We arrive at the same conclusion when comparing the graphs of the linear and the exponential functions, linear vs. exponential,  when $T_{max} = 100$ and $T_{max} = 25$. Figure \ref{distcomp} shows three pairs of functions when $T_{max} =$ 150, 100, and 25. As can be seen, in all three cases, the linear functions (the green curves) rises faster than the corresponding exponential functions (the purple curves), leading to more protected users over time.

\subsection{Experiment III: Collaborative Disinfection}
\label{exp3}
In this set of experiments, we examine the effects of \textit{collaborative disinfection}, which is defined in Section \ref{transinfimm}.   Collaborative disinfection allows infected users who are blocked by the malware from directly accessing AV provider web sites to get clean-up solutions from their OSN friends. 

We conducted on online survey asking 51 Facebook users if they would accept clean-up solutions from their Facebook friends if they were infected by such a malware.   Approximately 40\% (39.2\% to be exact) of the surveyed people responded said that they would do so and  would accept clean-up solutions from their Facebook friends. We use the result from this survey in this set of experiments, by setting $\delta_i$ to a maximum value of 0.4.  

\begin{figure}[t]
	\centering
	\subfigure[Set 1 - $\delta_i$=0.2,  $p_i$=0.5 ]
	{
		\includegraphics[width=2.5in]{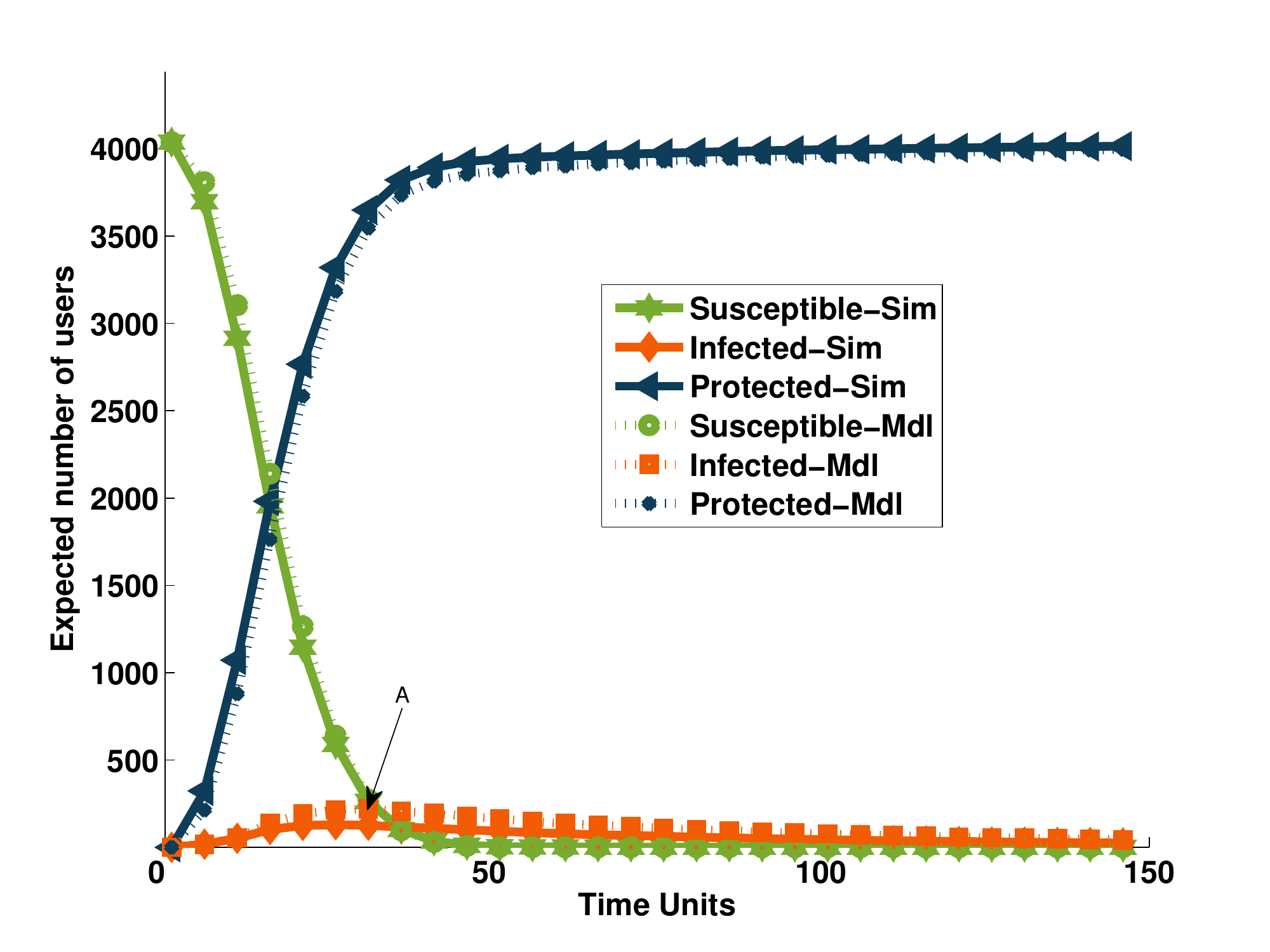}
		\label{exp3sc1}
	} 
	\subfigure[Set 1 - $\delta_i$=0.4, $p_i$=0.5 ]
	{
		\includegraphics[width=2.5in]{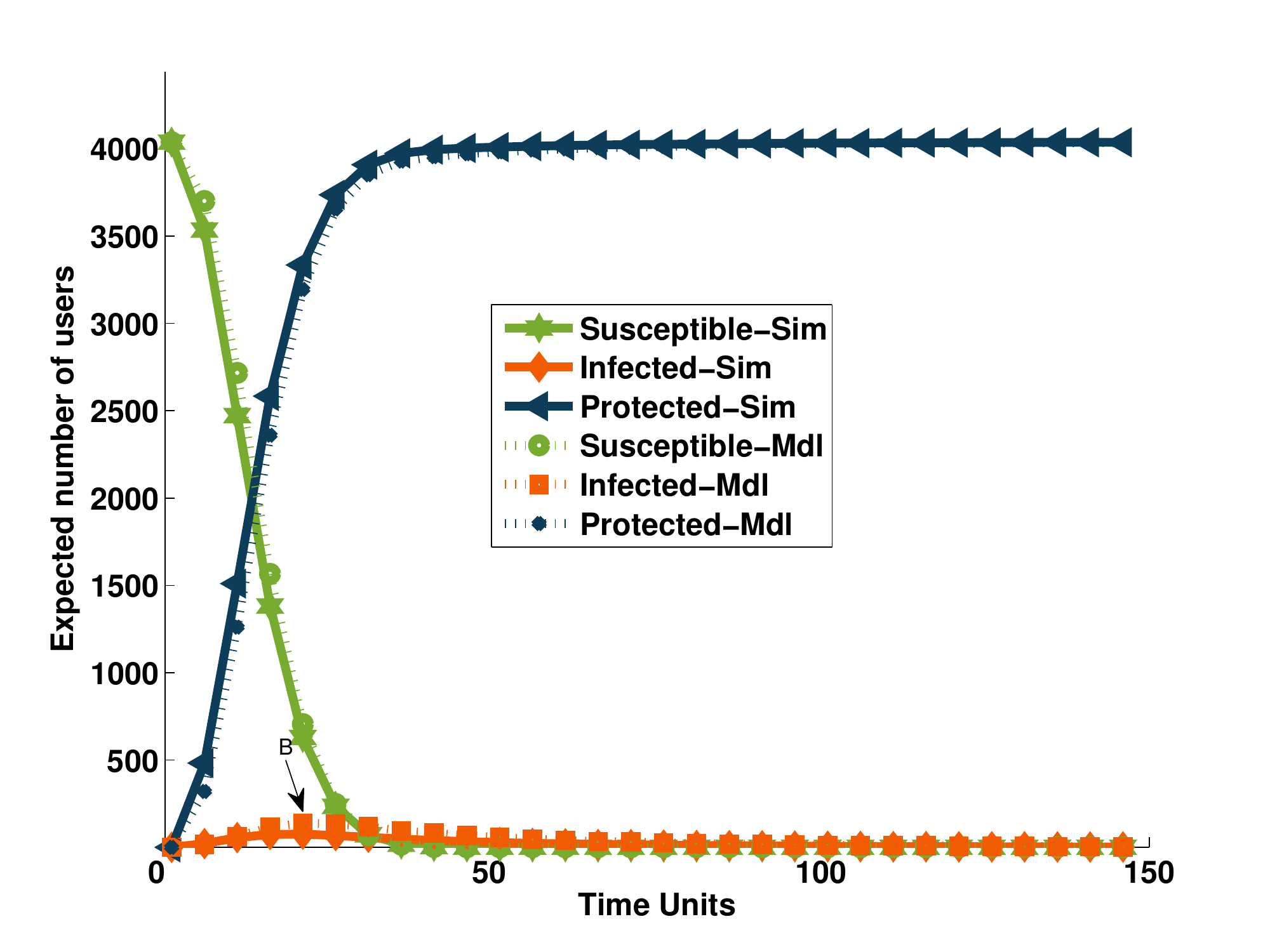}
		\label{exp3sc2}
	} 
	\subfigure[Set 1 - Number of infected users for different values of $\delta=$\{0,0.2,0.4\}   ]
	{
		\includegraphics[width=2.5in]{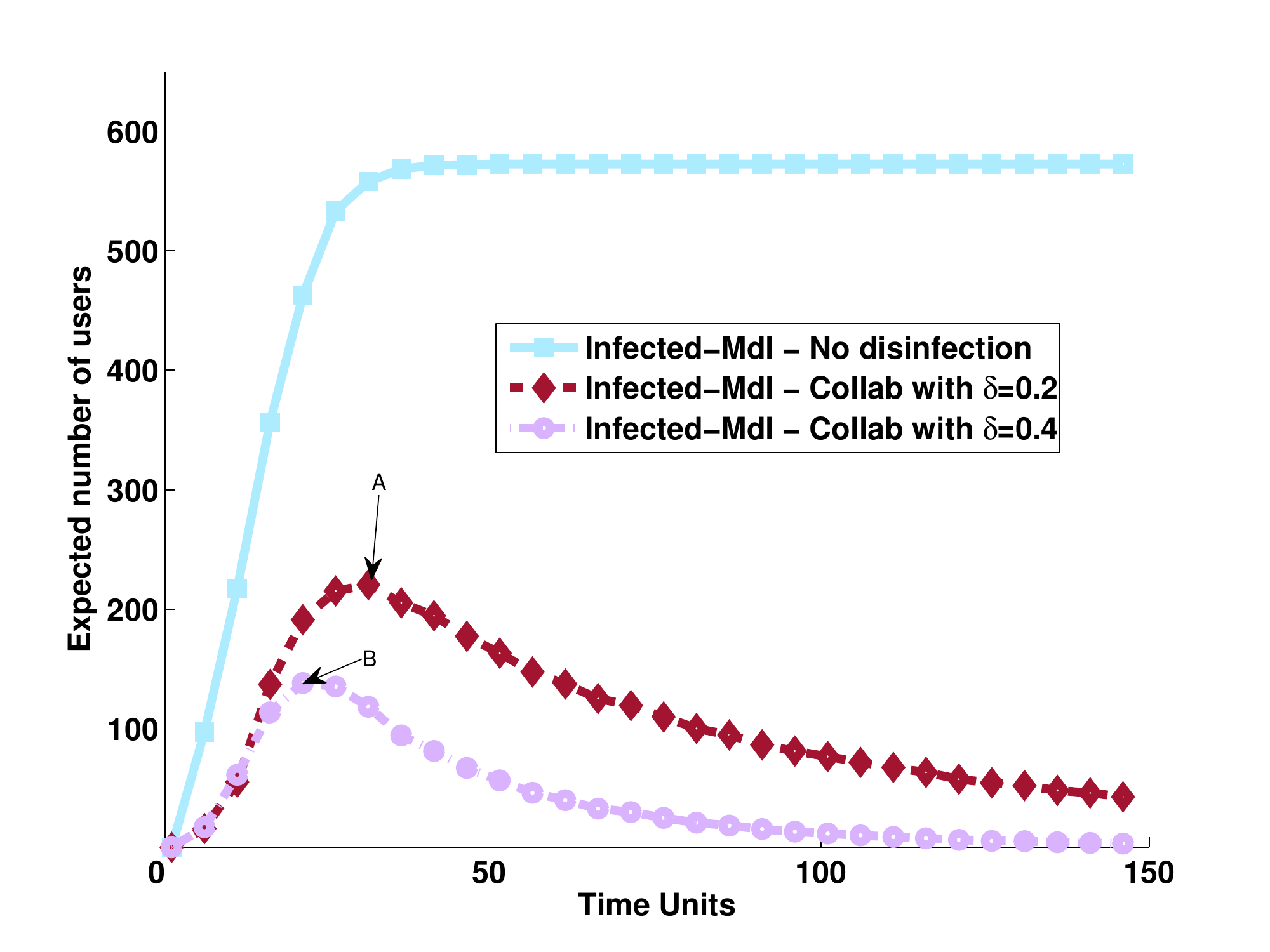}
		\label{exp3comp}
	}
	\caption{Experiment III: Collaborative disinfection -   $q_i$=0, $\hat{\beta}_{150}(t)= 0.005t$}
	\label{fig3}
\end{figure}

In this experiment, we examine different values of $\delta_i$, the probability of accepting clean-up solutions from friends.   Specifically, we consider $\delta_i = 0$, 0.2 and 0.4. (The value of 0.4 comes from the survey mentioned above.)  We assume that AV updates are released according to the linear function $\hat{\beta}_{150}(t)= 0.005t$,  in order to see the effectiveness of collaborative disinfection when comparing the results from this set of experiments with those obtained from Experiment II where no disinfection solutions were available.  As before, we assume no independent disinfection ($q_i = 0$) and users' probability of executing the malware is $p_i = 0.5$.   The results of this set of experiments are illustrated in Figures  \ref{exp2sc1}, \ref{exp3sc1} and \ref{exp3sc2} for $\delta_i = 0$, 0.2 and 0.4, respectively.

%
%
%
%
%

In the graphs, the number of infections first increases, then reaches a maximum value (points A and B in Fig. \ref{exp3sc1} and \ref{exp3sc2}, respectively).  After this point, the number of infections goes down thanks to infected users applying clean-up solutions suggested by their friends and moving from the infected state to the recovered  state.  As the number of infections going down, the number of protected users going up as users move from the infected state to the recovered state.

In general, collaborative disinfection plays an important role in containing (and almost stopping) the malware.  To illustrate this point, we extracted the curves of the number of infections from Fig. \ref{exp2sc1}, \ref{exp3sc2} and \ref{exp3sc2} and grouped them into one graph in Fig. \ref{exp3comp}.  The new graph shows that without clean-up solutions the number of infected users first increases then stays almost constant for the rest of the experiments because none of them is disinfected.   With collaborative disinfection, the number of infections goes down after reaching a maximum value (points A and B), thanks to clean-up solutions that allow user to be disinfected and move to the recovered state.  Furthermore, the higher the $\delta_i$ value, the more infected users transition to the recovered state.   In Fig. \ref{exp3comp},  there are 573 infected users in the network at the 50th time step in the case of no clean-up solutions, while this number reduces to 163 and 55 in the case of collaborative disinfection, when $\delta$ equals to 0.2 and 0.4, respectively.

Figures \ref{exp3sc1} and \ref{exp3sc2} also show that there is a good match between the analytical model and simulation results. For instance, in Fig. \ref{exp3sc1}, the discrepancy between model and simulation results are less than 6\% for all the susceptible, protected and infected graphs with the largest of 6\% at the twenty third step between the protected curves.  The Pearson correlation coefficient also shows close positive correlation between two series with at least $r \approx 0.98$ and $p-value \approx 0$ for all three curves,  susceptible, infected and protected. A similar comparison is also observed in Fig. \ref{exp3sc2} when $\delta_i=0.4$.

\subsection{Experiment IV: Independent Disinfection}
\label{exp4}
In this set of experiments, we examine the effects of \textit{independent disinfection} as discussed in Section \ref{transinfimm}.   Because the new malware prevents users from accessing directly AV provider web sites, knowledgeable users would search for clean-up solutions from third-party web sites not blocked by the malware. Another method is to access AV provider web sites via a second, clean computer and subsequently transfer the disinfecting software to the infected computer.  

We model these practices, called independent disinfection, using parameter $q_i$, where $q_i$ is the probability of user $i$ finding a clean-up solution without assistance from his/her OSN friends.

In this experiment, we consider different $q_i$ values, specifically, $q_i = 0$, 0.2 and 0.4.  We assume that AV updates are released according to the linear function $\hat{\beta}_{150}(t)= 0.005t$,  in order to see the effectiveness of collaborative disinfection when comparing the results from this set of experiments with those obtained from Experiment II where no disinfection solutions were available.  We assume no collaborative disinfection ($\delta_i = 0$).  Users' probability of executing the malware is $p_i = 0.5$.   

\begin{figure}
	\centering
	\subfigure[$q_i=0.2$]
	{
		\includegraphics[width=2.5in]{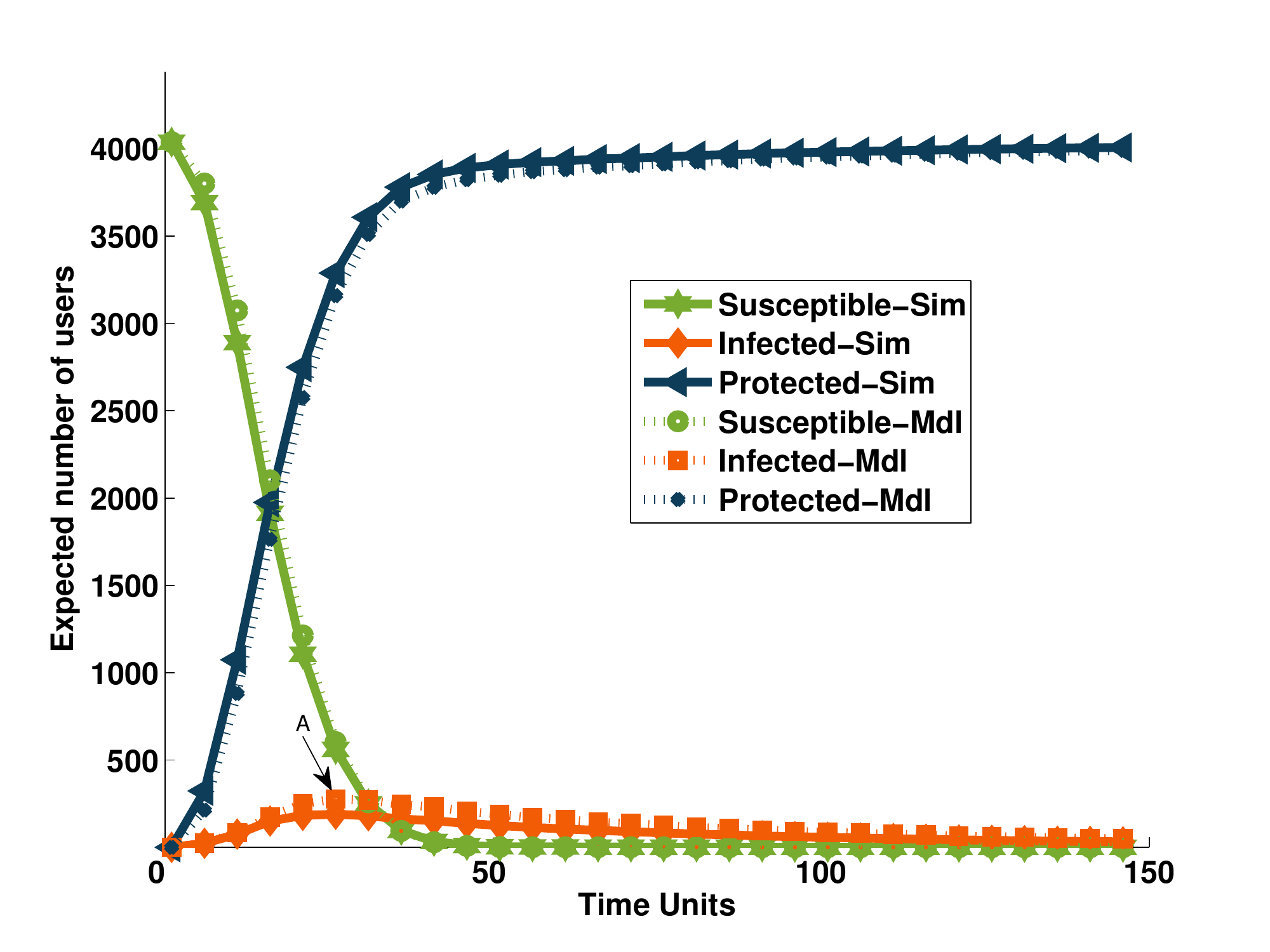}
		\label{exp4sc1}
	} 
	\subfigure[$q_i$=0.4]
	{
		\includegraphics[width=2.5in]{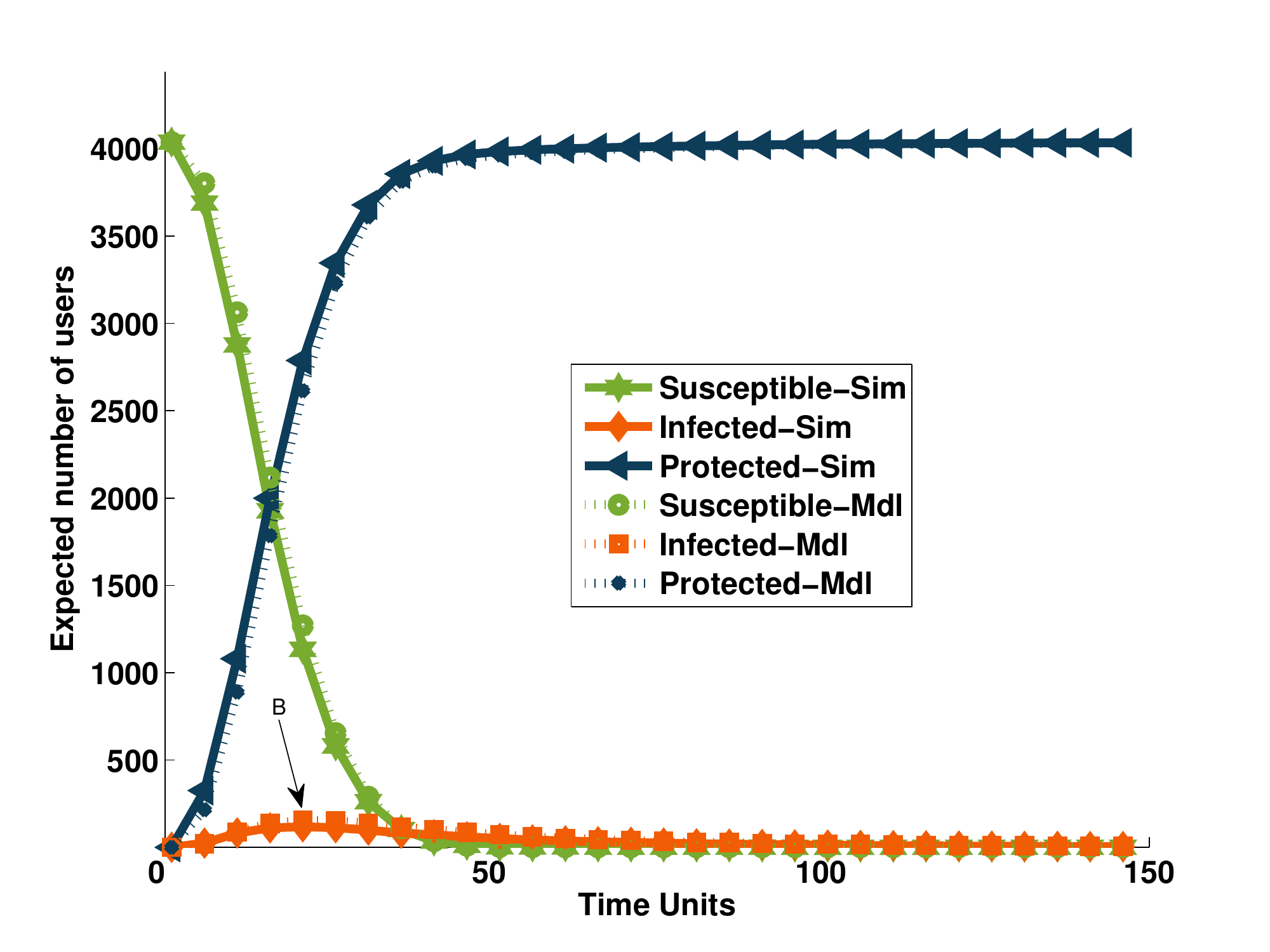}
		\label{exp4sc2}
	}
	\subfigure[Number of infected users for different values of $q=$\{0,0.2,0.4\}   ]
	{
		\includegraphics[width=2.5in]{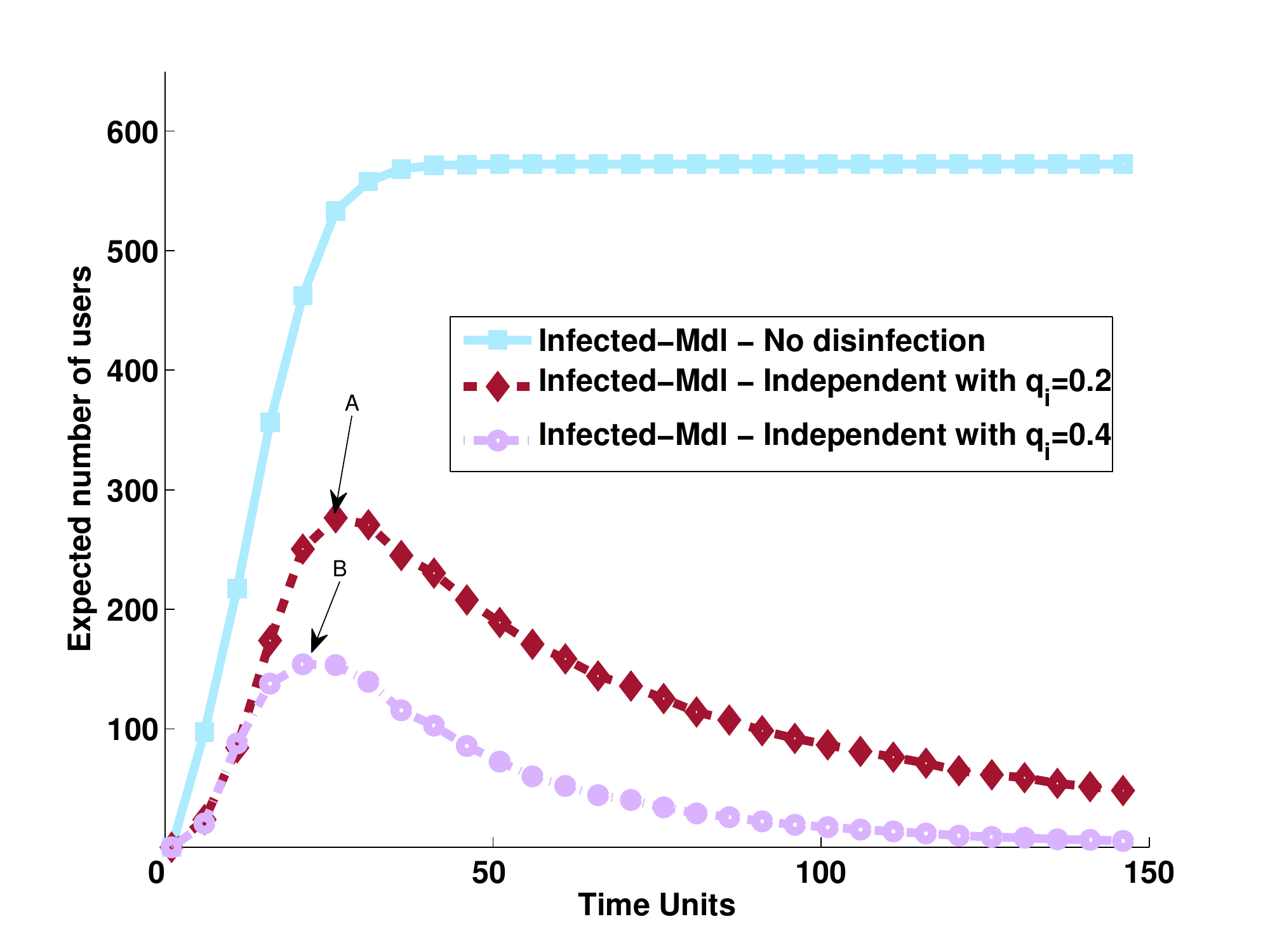}
		\label{exp4comp}
	}
	\caption{Experiment IV: Independent disinfection - $\delta_i$=0, $p_i$=0.5, $\hat{\beta}_{150}(t)= 0.005t$  }
	\label{fig4}
\end{figure}

The results of this set of experiments are illustrated in Fig.~\ref{exp4sc1} and \ref{exp4sc2} for $q_i = 0.2$ and 0.4, respectively.  As the graphs show, the number of infected users reaches a maximum value then goes down gradually until most  infected users become disinfected.  Clean-up solutions allow  users to disinfect themselves and move to the recovered (protected) state.  As the number of infected users decreases, the number of protected users rises. 

We also observe that higher $q_i$ values allow for better containment of the malware.  In other words,  as $q_i$ increases, the maximum number of infected users decreases. For example, in Fig. \ref{exp4sc1} when $q_i = 0.2$, the maximum number of infected users is equal to 277 (point A), while in Fig. \ref{exp4sc2}, this value for $q_i$=0.4 is equal to 151 (point B).   Fig.~\ref{exp4comp}, which combines the curves of the number of infected users from Fig. \ref{exp2sc1}, \ref{exp4sc1} and \ref{exp4sc2}, further illustrates this observation.  When $q_i = 0.4$, the number of infections goes down at a faster rate than that when $q_i = 0.2$.

Figure \ref{exp2sc1} depicts the case where $q_i = 0$ and the number of infected users increased and stayed constant until the end of the experiment because no disinfecting solutions were available.  In contrast, in Fig.~\ref{exp4sc1} and \ref{exp4sc2}, the number of infections decreases after points A and B, respectively, until most infected users are disinfected.  

In summary, knowledgeable users who find clean-up solutions independently are able to recover from the infected state, resulting in fewer infected users, and thus more protected users,  in the social network. The higher the value of $q_i$, the higher the number of protected users in the network.

\section{Experiment V: Frequency of Visits}
\label{exp5}

In the previous experiments, we assume that users visit the social network following an exponential distribution $\tau_i \sim E(40)$.  In this experiment, we assume that users visit the social network more often on average, following an exponential distribution $\tau_i \sim E(20)$.  

\begin{figure*}[tbh]
	\centering
	\subfigure[No disinfection - $p=0.5$, $\tau_i \sim E(20)$]
	{
		\includegraphics[width=2.5in]{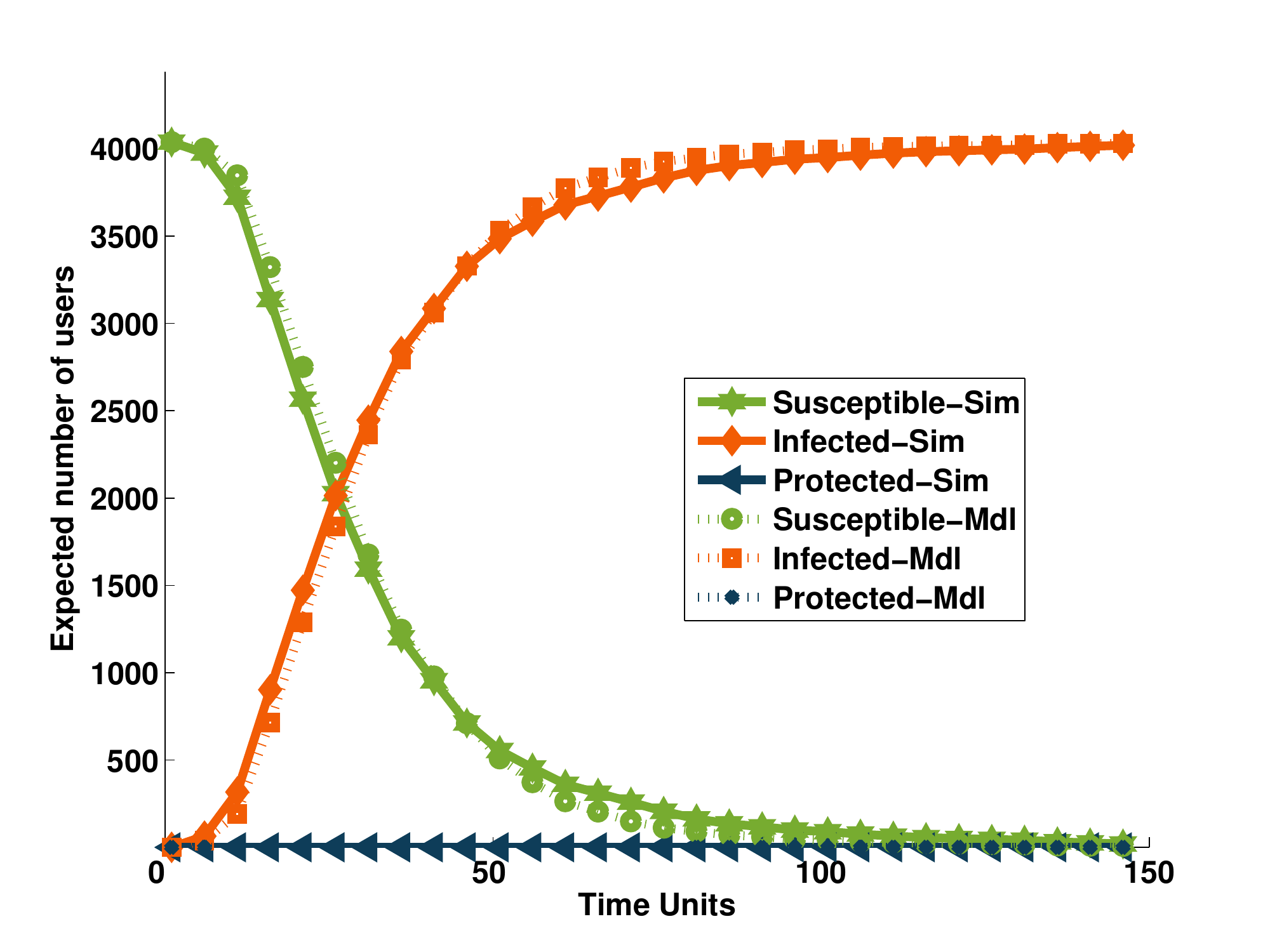}
		\label{exp5sc1}
	} 
	\subfigure[Collaborative disinfection -  $p=0.5$, $\hat{\beta}_{16}(t)= 0.05t$, $\delta=0.2$, $\tau_i \sim E(20)$ ]
	{
		\includegraphics[width=2.5in]{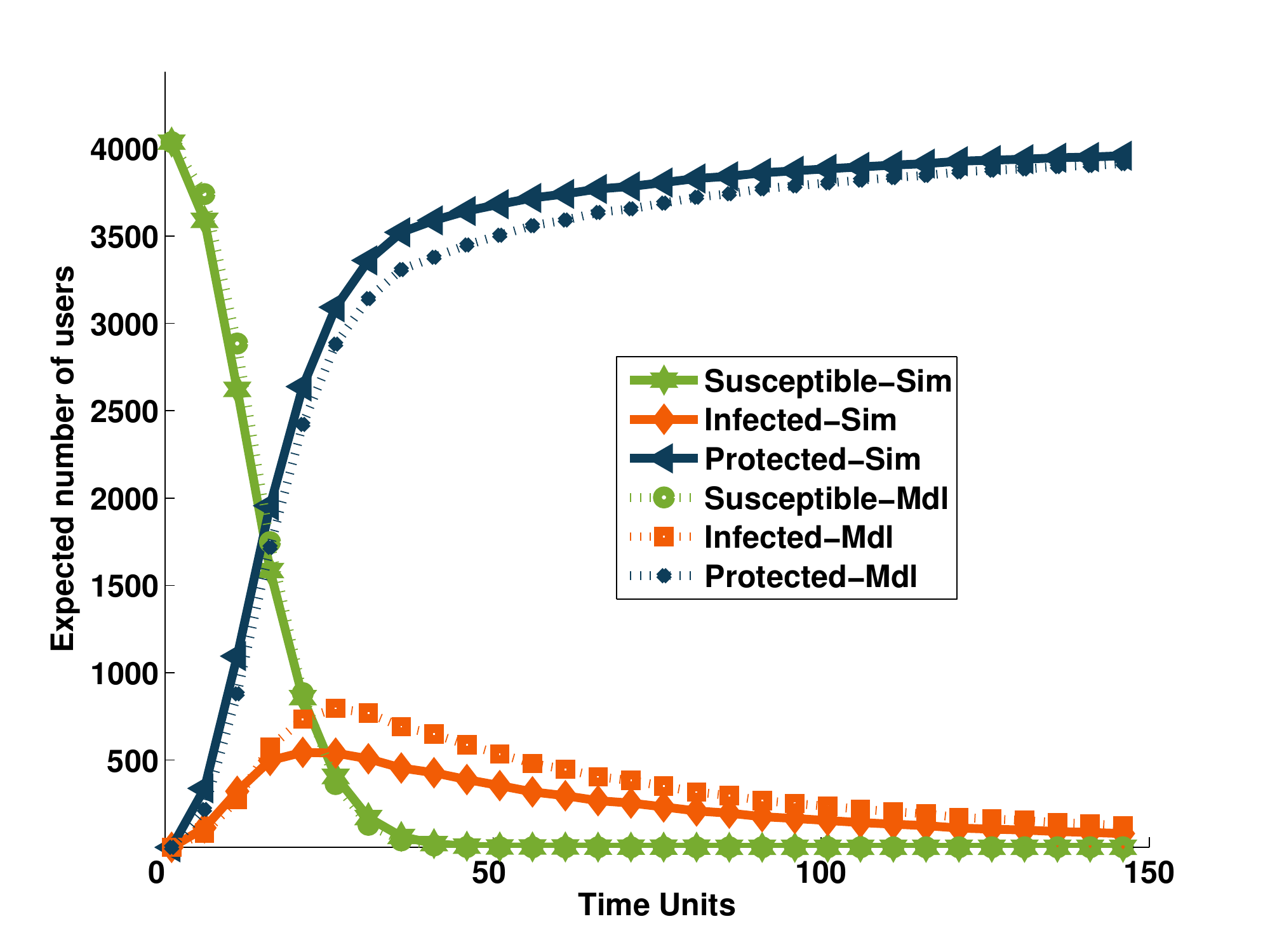}
		\label{exp5sc2}
	}
	\subfigure[No disinfection - Number of infected users with $\tau_i \sim E(20)$ vs. $\tau_i \sim E(40)$]
	{
		\includegraphics[width=2.5in]{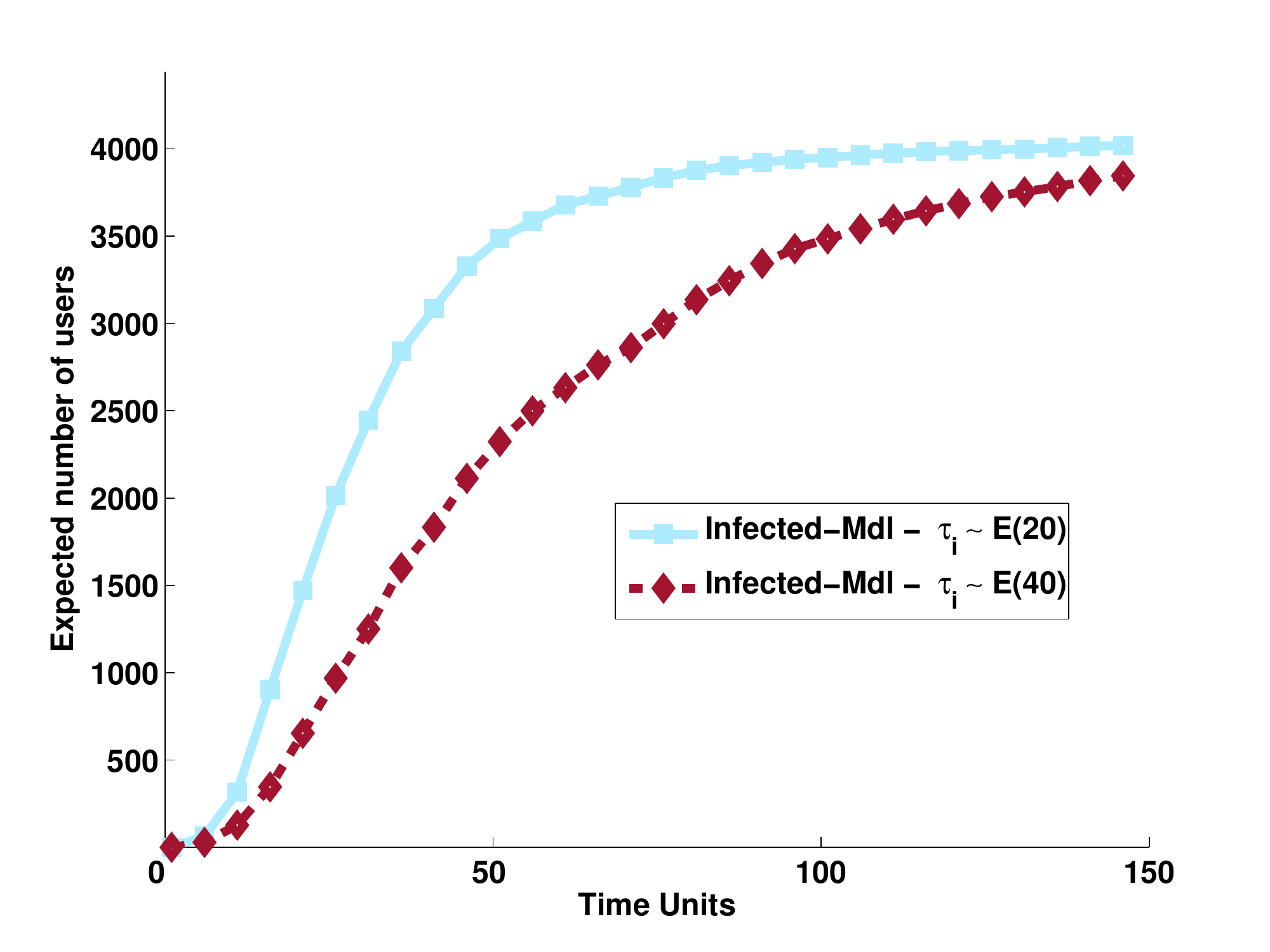}
		\label{exp5sc3}
	}
	\subfigure[Collaborative disinfection -  Number of infected users with $\tau_i \sim E(20)$ vs. $\tau_i \sim E(40)$]
	{
		\includegraphics[width=2.5in]{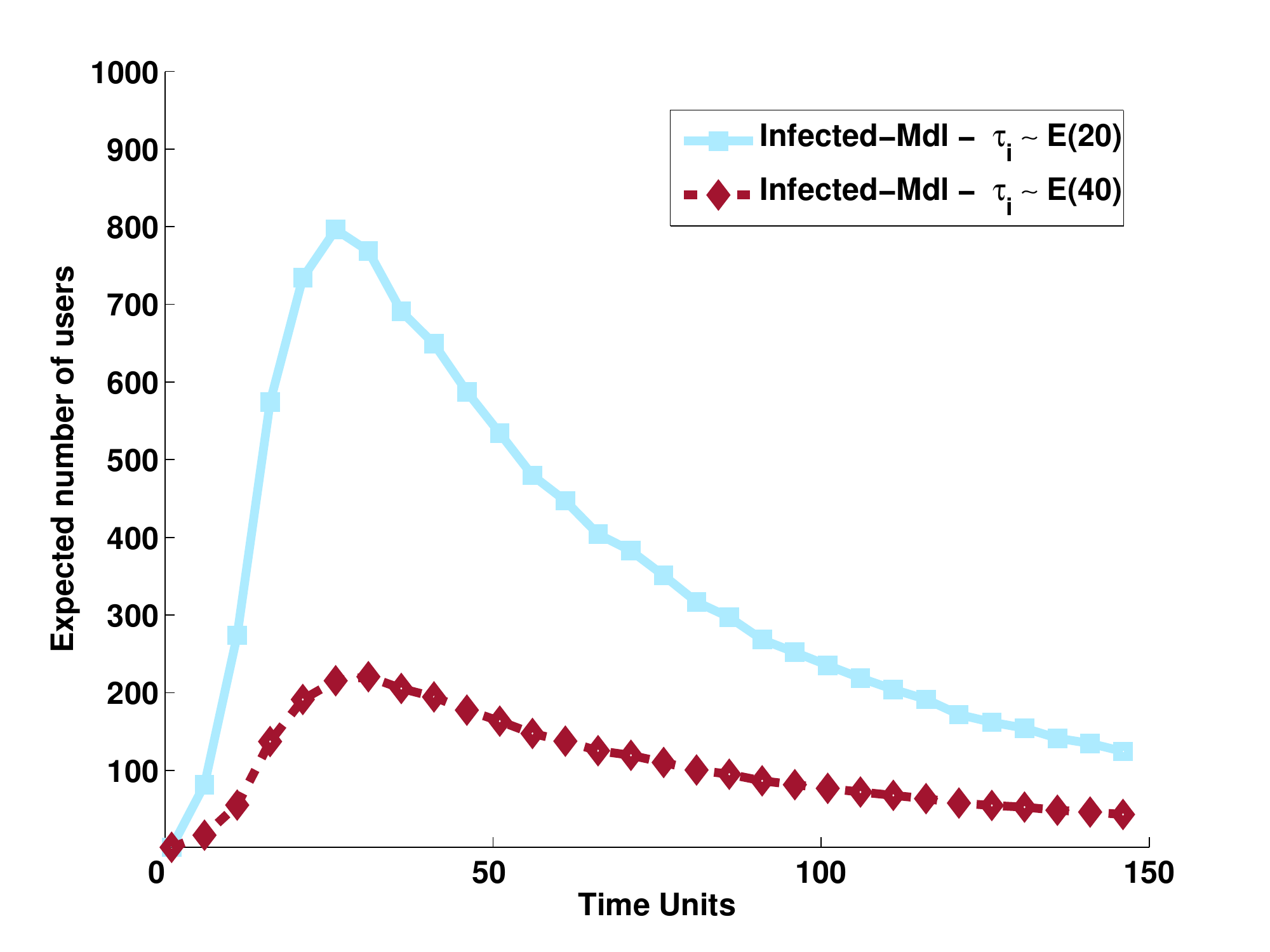}
		\label{exp5sc4}
	}
	\caption{Experiment V: Frequency of visits  - $\tau \sim E(20)$ vs. $E(40)$  }
	\label{fig5}
\end{figure*}

We repeated the experiments whose results are shown in Figs. \ref{exp1sc1} and \ref{exp3sc1} using the new user message checking time $\tau_i \sim E(20)$.  The new experimental results are given in Figs \ref{exp5sc1} and \ref{exp5sc2}, respectively.  The new graphs show the same trends as those in the previous experiments.  When there is no AV update and no clean-up solution (Fig. \ref{exp5sc1}), the number of infected users increases over time.  With gradual linear AV release updates and collaborative disinfection (Fig. \ref{exp5sc2}), the number of infected users increases at first, until it reaches to a certain point. After that,  the number of infected users goes down thanks to collaborative disinfection.  

In order to compare the impact of frequency of visits, we consolidated the curves representing the numbers of infected users from Fig. \ref{exp1sc1} (for $\tau_i \sim E(40)$) and Fig. \ref{exp5sc1} (for $\tau_i \sim E(20)$) and placed them in Fig. \ref{exp5sc3}.   The combined graph shows that the higher the frequency of visits, the higher the number of infected users within the same time frame.  For instance, at the 50th time unit, there are 2,323 infected users in the system when $\tau_i \sim E(40)$, while this number is 3,484 when $\tau_i \sim E(20)$, a staggering difference of 30\% more infected users within the same time frame.

We also combined the curves representing the numbers of infected users from Fig. \ref{exp3sc1} (for $\tau_i \sim E(40)$) and Fig. \ref{exp5sc2} (for $\tau_i \sim E(20)$) and placed them in Fig. \ref{exp5sc4}.   We observe a similar comparison:  the higher the frequency of visits, the higher the number of infected users within the same time frame.  For instance, at the 25th time unit, there are 793 infected users in the network when $\tau_i \sim E(20)$. This number is 215 infected users when $\tau_i \sim E(40)$.
	

\section{Related Work}
\label{relworks}
We broadly classify existing research related to our work  into two topics: (1) studies and modeling of malware propagation dynamics, and (2) countermeasures.

\subsection{Studies and Modeling of Malware Propagation Dynamics}
Most existing studies on propagations of Trojans in OSNs are based on {\em simulations} \cite{dyn, ftrojan, fworm, facebook-virus} or {\em experiments}~\cite{koobface}.  
Yan et al. \cite{dyn} studied the impact of network topology and users' probability of following malicious links on the propagation of Trojans (which are termed ``active worms'' in \cite{dyn}).  Their simulation results suggest that infected users who seldom visit the social network (i.e., equivalent to being quarantined)  help to slow down the propagation of worms.  On the other hand, the higher the probability that users follow malicious links, the faster a worm spreads in the network.  
Faghani and his collaborators \cite{ftrojan, fworm} also showed via simulations that the higher the probability that users click on malicious links,  the faster a malware propagates in a social network.
Fan et al. \cite{facebook-virus} studied propagations of worms/viruses hidden in (malicious) third-party applications built on the Facebook application platform.  Their simulation results show that worms/viruses spread faster on Facebook than in an email network due to (1) users spending more time on Facebook than on email, and (2) the topological characteristics of OSNs, namely, power-law distribution of node degrees and high clustering coefficient.  
Thomas et al. \cite{koobface} traced the activities of Koobface, a Trojan that targeted OSN users, for one month to study its propagation characteristics. 

Existing works on modelling malware propagation in {\em online social networks} include \cite{fxssj, fxss} and \cite{twitjack}.  Faghani and his collaborators  \cite{fxssj, fxss} modeled the propagation of {\em cross-site-scripting} (XSS) worms in OSNs.   Sanzgiri et al. \cite{twitjack} modeled the propagation of Trojans in the social network Twitter, which is represented by {\em directed graphs} due to one-directional (follower-followee) relationships.   To the best of our knowledge,  the analytical model we propose in this article is the first that characterizes the propagation of Trojans in social networks represented by {\em undirected graphs} such as Facebook, LinkedIn and Orkut.

There exist works that model the propagation of worms and malware (not necessarily Trojans) in other types of networks such as people, email and cellular phones.  Many of these models \cite{boguna,moreno,moreno2,pastor} assumed that each user is directly connected to every other user in the same network (also known as ``homogeneous mixing'').  This assumption does not hold true for a real-world OSN such as Facebook where each user is directly connected to only his/her friends.  As a result, the ``homogeneous mixing'' assumption  may lead to an over-estimation of the infection rate in a real OSN~\cite{wen,email}.  Cheng et al. \cite{smartchen} proposed a propagation model for malware that targets multimedia messaging service (MMS) and bluetooth devices; this model also assumes ``homogeneous mixing'', and is thus not applicable to real OSNs.   Chen and Ji \cite{chen} and Chen et al. \cite{sii1} modeled the spreading of scanning worms in computer networks.  Zou et al \cite{email} proposed a propagation model for email worms that takes into account user behavior such as checking time and probability of opening email attachments.  Komnios et al. \cite{komin} modeled the propagation of email worms using differential equations.  Wen et al. \cite{wen} also modeled the propagation of malware in email networks and in semi-directed networks represented by mixed graphs (i.e., a subset of edges are directed while the others are undirected).

\subsection{Countermeasures}
Detecting malware is also an active and important research topic in social networks. For instance, Rahman et al. \cite{socware} presented a Facebook application that identifies \quotes{socware} in OSNs. They defined socware as any parasitic behavior in OSNs. This includes posts that spread malware, web pages pointed to malware, false reward posts, rogue Facebook applications, requests, surveys, and likejacking. They used machine learning techniques to distinguish between socware posts and benign posts. They estimated the false negative rate of their system to be about 0.3\%.

Xu et al. \cite{xu} proposed a correlation-based scheme to detect active worm propagation in OSNs. They assigned \quotes{decoy friends} to a subset of users, and the \quotes{decoy friends} monitored network activities. 
Stringhini et al. \cite{fbspam} studied the detection of spammers using \quotes{honey} profiles in three major OSNs: Facebook, Twitter and MySpace. They applied their classifier to about 800,000 profiles and detected 130 spammers, most of which were related to adult websites.

Yan et al. \cite{dyn} described three approaches for monitoring users in OSNs in order to detect malware using (1) node degree metric, (2) user activities and (3) network partition into small islands.  In the first approach, nodes with the highest degrees are chosen to be monitored.  In the second approach, the most active nodes are selected for monitoring.  Examples of the most active nodes are major broadcasting companies such as CNN and BBC, which post news updates frequently throughout the day via OSNs (e.g., Facebook and Twitter networks).  In the third approach, an OSN is partitioned into small islands, and every message exchanged between islands is inspected for potential viruses/malware. 

Faghani and Nguyen \cite{fxssj} also studied the effectiveness of several metrics used for monitoring user activities in OSNs to detect malware.    Their simulation results show that monitoring users that have high numbers of connections to different communities (i.e., high cross-clique connectivity) help to detect propagating malware earlier than using the other metrics. 


Cao et al. presented PathCutter \cite{pathcutter}, a tool that can detect traditional DOM-based XSS, and content sniffing XSS vulnerabilities in social networks. 

Livshits et al.~\cite{spectator} designed a detection and containment method called Spectator \cite{spectator}. Spectator uses a tainting and tagging approach to detect the spread of JavaScript worms which is implemented as a proxy. 


\section{Conclusion}
\label{concl}
In this article, we present an analytical model to study  propagation characteristics of Trojan malware and  factors that impact the propagation dynamics of Trojans in an online social network.

Unlike most previous works, the proposed model assumes all the topological characteristics of real online social networks, namely, low average shortest distance, power-law distribution of node degrees and high clustering coefficient.  
Furthermore, the model takes into account attacking trends of modern Trojans (e.g., their ability to block users' access to AV provider websites),  the role of AV products, and security practices such as gradual AV update release by AV providers and users' collaborative disinfection.  These factors were never considered in existing works.  By taking into account these factors, the proposed model can accurately and realistically estimate the infection rate caused by a Trojan malware in an OSN as well the recovery rate of the user population.

The model is validated using a Facebook sub-graph.  The numerical results obtained from the model closely match the simulation results.   While being accurate, the model also has low computational complexity, in the order of $O(E)$, where $E$ is the number of edges in the network graph.  

From the numerical and simulation results, we draw the following conclusions and lessons.
AV products play an important role in protecting OSN users from Trojans.
For zero-day or very novel malware, the faster AV providers release updates/patches, the more users will be protected.
In the case of  blocking malware, collaborative disinfection is an effective mechanism that helps infected users to recover, especially in cases of sophisticated Trojans that use advanced social engineering techniques to deceive OSN users.

User awareness of security threats and safe browsing practices play an important role in protecting OSN users from Trojans by slowing down the propagation of malware.  OSN administrators should launch campaigns and advertisements to educate users about safe browsing practices (e.g., not following unknown links) and about new malicious social engineering techniques as soon they are discovered. In the case of blocking malware, OSN providers should notify infected users via different channels, e.g., short message service (SMS) or email, and provide them with clean-up solutions as early as possible.

%

In our future work, we will enhance our model to include multiple infiltrating nodes.  We will conduct surveys to accurately model user behaviour of following unknown links and executing hidden malicious code (probability $p_i$).  We will also research timelines of AV updates released by AV providers in past attacks to derive different functions $\hat{\beta}(t)$ that reflect real-world practices.

\ifCLASSOPTIONcaptionsoff
  \newpage
\fi

\bibliographystyle{ieeetr} 
\bibliography{refscleaned}

\begin{thebibliography}{10}

\bibitem{alexa}
Alexa, ``The top 500 sites on the web.'' \url{https://goo.gl/E2GtDp}, December
  2014.

\bibitem{grossman}
J.~Grossman, ``Cross-site scripting worms and viruses: the impending threat and
  the best defense,'' June 2007.

\bibitem{facebook-virus}
W.~Fan and K.~Yeung, ``Online social networks paradise of computer viruses,''
  {\em Journal of Physica A: Statistical Mechanics and its Applications},
  vol.~390, pp.~189 -- 197, January 2011.

\bibitem{opin}
K.~Thomas, C.~Grier, and V.~Paxson, ``Adapting social spam infrastructure for
  political censorship,'' in {\em Proceedings of the 5th USENIX Workshop on
  Large-Scale Exploits and Emergent Threats}, (Berkeley, CA), USENIX, April
  2012.

\bibitem{designsocialbot}
Y.~Boshmaf, I.~Muslukhov, K.~Beznosov, and M.~Ripeanu, ``Design and analysis of
  a social botnet,'' {\em Elsevier Journal on Computer Networks}, vol.~57,
  pp.~556--578, February 2013.

\bibitem{cj}
R.~Hansen and J.~Grossman, ``Clickjacking,'' September 2008.

\bibitem{extmalware}
B.~Li, ``An in-depth look into malicious browser extensions.''
  \url{https://goo.gl/r6lwO4}, October 2014.

\bibitem{kilim1}
ESET, ``Virus radar.'' \url {https://goo.gl/2bpwsB}, November 2014.

\bibitem{kasper}
{Kaspersky Labs}, ``Facebook malware poses as flash update, infects 110k
  users,'' February 2015.

\bibitem{koobface}
K.~Thomas and D.~Nicol, ``The {Koobface} botnet and the rise of social
  malware,'' in {\em Proceedings of the 5th International Conference on
  Malicious and Unwanted Software (MALWARE 2010)}, pp.~63--70, October 2010.

\bibitem{kilim2}
ESET, ``Virus radar.'' \url {https://goo.gl/LVDEJS}, November 2016.

\bibitem{checkpoint}
R.~Ziakin and D.~Barda, ``Imagegate: Check point uncovers a new method for
  distributing malware through images.'' \url{https://goo.gl/MiU0jU}, November
  2016.

\bibitem{dyn}
G.~Yan, G.~Chen, S.~Eidenbenz, and N.~Li, ``Malware propagation in online
  social networks: Nature, dynamics, and defense implications,'' in {\em
  Proceedings of the 6th ACM Symposium on Information, Computer and
  Communications Security}, ASIACCS '11, (New York, NY, USA), pp.~196--206,
  ACM, March 2011.

\bibitem{ftrojan}
M.~Faghani, A.~Matrawy, and C.-H. Lung, ``A study of trojan propagation in
  online social networks,'' in {\em In Proceedings of the 5th International
  Conference on New Technologies, Mobility and Security (NTMS)}, pp.~1--5, May
  2012.

\bibitem{fworm}
M.~Faghani and H.~Saidi, ``Malware propagation in online social networks,'' in
  {\em Proceedings of the International Conference on Malicious and Unwanted
  Software (MALWARE)}, pp.~8--14, October 2009.

\bibitem{fxssj}
M.~Faghani and U.~T. Nguyen, ``A study of xss worm propagation and detection
  mechanisms in online social networks,'' {\em IEEE Transactions Journal on
  Information Forensics and Security}, vol.~8, pp.~1815--1826, November 2013.

\bibitem{fxss}
M.~Faghani and H.~Saidi, ``Social networks' xss worms,'' in {\em Proceedings of
  International Conference on Computational Science and Engineering}, vol.~4,
  pp.~1137--1141, August 2009.

\bibitem{twitjack}
A.~Sanzgiri, J.~Joyce, and S.~Upadhyaya, ``The early (tweet-ing) bird spreads
  the worm: An assessment of twitter for malware propagation,'' {\em Procedia
  Computer Science Journal}, vol.~10, pp.~705 -- 712, August 2012.

\bibitem{boguna}
M.~Bogun{\'a}, R.~Pastor-Satorras, and A.~Vespignani, {\em Epidemic spreading
  in complex networks with degree correlations}, pp.~127--147.
\newblock Berlin, Heidelberg: Springer Berlin Heidelberg, September 2003.

\bibitem{moreno}
Y.~Moreno, J.~B. G\'omez, and A.~F. Pacheco, ``Epidemic incidence in correlated
  complex networks,'' {\em Physical Review E Journal}, vol.~68, p.~035103, Sep
  2003.

\bibitem{moreno2}
Y.~Moreno, R.~Pastor-Satorras, and A.~Vespignani, ``Epidemic outbreaks in
  complex heterogeneous networks,'' {\em The European Physical Journal
  B-Condensed Matter and Complex Systems}, vol.~26, pp.~521--529, July 2002.

\bibitem{pastor}
R.~Pastor-Satorras and A.~Vespignani, ``Epidemic spreading in scale-free
  networks,'' {\em Physical Review Letters}, vol.~86, pp.~3200--3203, April
  2001.

\bibitem{wen}
S.~Wen, W.~Zhou, J.~Zhang, Y.~Xiang, W.~Zhou, and W.~Jia, ``Modeling
  propagation dynamics of social network worms,'' {\em IEEE Transactions
  Journal on Parallel and Distributed Systems}, vol.~24, pp.~1633--1643, August
  2013.

\bibitem{email}
C.~Zou, D.~Towsley, and W.~Gong, ``Modeling and simulation study of the
  propagation and defense of internet e-mail worms,'' {\em IEEE Transactions
  Journal on Dependable and Secure Computing}, vol.~4, pp.~105--118, April
  2007.

\bibitem{smartchen}
S.-M. Cheng, W.~C. Ao, P.-Y. Chen, and K.-C. Chen, ``On modeling malware
  propagation in generalized social networks,'' {\em IEEE Communications
  Letters Journal}, vol.~15, pp.~25--27, January 2011.

\bibitem{chen}
Z.~Chen and C.~Ji, ``Spatial-temporal modeling of malware propagation in
  networks,'' {\em IEEE Transactions Journal on Neural Networks}, vol.~16,
  pp.~1291--1303, September 2005.

\bibitem{sii1}
Z.~Chen, L.~Gao, and K.~Kwiat, ``Modeling the spread of active worms,'' in {\em
  Proceedings of Twenty-Second Annual Joint Conference of the IEEE Computer and
  Communications. IEEE Societies (INFOCOM 2003)}, vol.~3, pp.~1890--1900, IEEE,
  March 2003.

\bibitem{komin}
T.~Komninos, Y.~C. Stamatiou, and G.~Vavitsas, {\em A worm propagation model
  based on people's email acquaintance profiles}, pp.~343--352.
\newblock Berlin, Heidelberg: Springer Berlin Heidelberg, December 2006.

\bibitem{avnews}
S.~Sjouwerman, ``Bad news: Your antivirus detection rates have dramatically
  declined in 12 months.'' \url {https://goo.gl/JTa5jo}, January 2017.

\bibitem{conficker}
P.~Porras, H.~Saidi, and V.~Yegneswaran, ``Conficker {C} analysis,'' tech.
  rep., SRI International, June 2009.

\bibitem{sophoslab}
{Sophos Labs}, ``Virus threat protection.'' \url{https://goo.gl/EEkDha},
  December 2015.

\bibitem{ghost}
N.~Provos, D.~McNamee, P.~Mavrommatis, K.~Wang, and N.~Modadugu, ``The ghost in
  the browser analysis of web-based malware,'' in {\em Proceedings of the First
  Conference on First Workshop on Hot Topics in Understanding Botnets},
  HotBots'07, (Berkeley, CA, USA), pp.~4--4, USENIX Association, October 2007.

\bibitem{dekker}
A.~Dekker, ``Realistic social networks for simulation using network rewiring,''
  in {\em Proceedings of the International Congress on Modelling and
  Simulation}, pp.~677--683, May 2007.

\bibitem{holme}
P.~Holme and B.~J. Kim, ``Growing scale-free networks with tunable
  clustering,'' {\em Journal of Physical Review E}, vol.~65, p.~026107,
  September 2002.

\bibitem{mislov}
A.~Mislove, M.~Marcon, K.~P. Gummadi, P.~Druschel, and B.~Bhattacharjee,
  ``Measurement and analysis of online social networks,'' in {\em Proceedings
  of the 7th ACM SIGCOMM Conference on Internet Measurement}, IMC '07, (New
  York, NY, USA), pp.~29--42, ACM, October 2007.

\bibitem{snap}
J.~Mcauley and J.~Leskovec, ``Discovering social circles in ego networks,''
  {\em ACM Transaction Journal on Knowledge Discovery Data}, vol.~8,
  pp.~4:1--4:28, February 2014.

\bibitem{nman}
M.~J. Newman, ``Power laws, pareto distributions and zipf's law,'' {\em Journal
  on Contemporary {Physics}}, vol.~46, pp.~323--351, November 2005.

\bibitem{dis}
Y.~Zhou and X.~Jiang, ``Dissecting android malware: Characterization and
  evolution,'' in {\em Proceedings of the 2012 IEEE Symposium on Security and
  Privacy (SP)}, pp.~95--109, May 2012.

\bibitem{p2p}
R.~Thommes and M.~Coates, ``Epidemiological modelling of peer-to-peer viruses
  and pollution,'' in {\em Proceedings of the 25th IEEE International
  Conference on Computer Communications}, pp.~1--12, April 2006.

\bibitem{sis1}
A.~Ganesh, L.~Massoulie, and D.~Towsley, ``The effect of network topology on
  the spread of epidemics,'' in {\em Proceedings of 24th IEEE Annual Joint
  Conference of the IEEE Computer and Communications Societies}, vol.~2,
  pp.~1455--1466, March 2005.

\bibitem{sis2}
Y.~Wang, D.~Chakrabarti, C.~Wang, and C.~Faloutsos, ``Epidemic spreading in
  real networks: an eigenvalue viewpoint,'' in {\em Proceedingss of 22nd
  International Symposium on Reliable Distributed Systems}, pp.~25--34, October
  2003.

\bibitem{Microsoft}
D.~Batchelder, J.~Blackbird, and P.~Henry, ``{Microsoft Security Intelligence
  Report},'' tech. rep., Microsoft, June 2014.

\bibitem{evas1}
J.~Marpaung, M.~Sain, and H.-J. Lee, ``Survey on malware evasion techniques:
  State of the art and challenges,'' in {\em Proceedings of the 14th
  International Conference on Advanced Communication Technology (ICACT)},
  pp.~744--749, February 2012.

\bibitem{evas2}
N.~Thamsirarak, T.~Seethongchuen, and P.~Ratanaworabhan, ``A case for malware
  that make antivirus irrelevant,'' in {\em Proceedings of the 12th
  International Conference on Electrical Engineering/Electronics, Computer,
  Telecommunications and Information Technology (ECTI-CON)}, pp.~1--6, June
  2015.

\bibitem{sigheu}
Y.~Ye, D.~Wang, T.~Li, D.~Ye, and Q.~Jiang, ``An intelligent {PE-malware}
  detection system based on association mining,'' {\em Journal in Computer
  Virology}, vol.~4, pp.~323--334, October 2008.

\bibitem{avtest}
{AV-Test Lab}, ``{Anti-Virus} outbreak response testing and impact.'' \url
  {https://goo.gl/sNaEiX}, January 2004.

\bibitem{MMPConficker}
Microsoft, ``Conficker worm: Help protect windows from {Conficker}.''
  \url{https://goo.gl/3T3Du8}, February 2010.

\bibitem{trendconfick}
{Sophos Labs}, ``{Mal/Conficker-A}.'' \url {https://goo.gl/yQiFnx}, October
  2011.

\bibitem{sophos}
F.~Howard and O.~Komili, ``Poisoned search results: How hackers have automated
  search engine poisoning attacks to distribute malware,'' tech. rep., Sophos
  Labs, 2010.

\bibitem{fakeAV}
B.~Stone-Gross, R.~Abman, R.~A. Kemmerer, C.~Kruegel, D.~G. Steigerwald, and
  G.~Vigna, ``The underground economy of fake antivirus software,'' in {\em
  Economics of Information Security and Privacy III}, pp.~55--78, Springer,
  July 2013.

\bibitem{pearson}
K.~Pearson, ``Note on regression and inheritance in the case of two parents,''
  {\em Journal of the Royal Society of London}, pp.~240--242, January 1895.

\bibitem{microreport}
C.~Anthe, P.~Charzan, and E.~Florio, ``Microsoft security intelligence
  report,'' April 2015.

\bibitem{socware}
M.~S. Rahman, T.-K. Huang, H.~V. Madhyastha, and M.~Faloutsos, ``Efficient and
  scalable socware detection in online social networks,'' in {\em Proceedings
  of the 21st USENIX Conference on Security Symposium}, Security'12, (Berkeley,
  CA, USA), pp.~32--32, USENIX Association, August 2012.

\bibitem{xu}
W.~Xu, F.~Zhang, and S.~Zhu, ``Toward worm detection in online social
  networks,'' in {\em Proceedings of the 26th Annual Computer Security
  Applications Conference}, ACSAC '10, (New York, NY, USA), pp.~11--20, ACM,
  2010.

\bibitem{fbspam}
G.~Stringhini, C.~Kruegel, and G.~Vigna, ``Detecting spammers on social
  networks,'' in {\em Proceedings of the 26th Annual Computer Security
  Applications Conference}, ACSAC '10, (New York, NY, USA), pp.~1--9, ACM,
  December 2010.

\bibitem{pathcutter}
Y.~Cao, V.~Yegneswaran, P.~A. Porras, and Y.~Chen, ``Pathcutter: Severing the
  self-propagation path of xss javascript worms in social web networks,'' in
  {\em Proceedings of NDSS Conference}, The Internet Society, February 2012.

\bibitem{spectator}
V.~B. Livshits and W.~Cui, ``Spectator: Detection and containment of javascript
  worms.,'' in {\em Proceedings of USENIX Annual Technical Conference}
  (R.~Isaacs and Y.~Zhou, eds.), pp.~335--348, USENIX Association, July 2008.

\end{thebibliography}

\vfill
%
%
%




\end{document}